\documentclass[11pt]{article}
\usepackage[T1]{fontenc}
\usepackage[utf8]{inputenc}
\usepackage[left=2.75cm, right=2.75cm, top=3cm, bottom=3cm]{geometry}
\usepackage{tabu, makecell, array, colortbl, booktabs, multirow}
	\newcolumntype{C}[1]{>{\centering\arraybackslash}p{#1}}

\usepackage[format=plain, textfont=it, labelfont={it, bf}]{caption}

\usepackage{graphicx, subcaption}

\usepackage{authblk}

\usepackage{natbib}

\usepackage{hyperref}
	\hypersetup{
		colorlinks,
		citecolor=black,
		filecolor=black,
		linkcolor=black,
		urlcolor=black
	}
	
\usepackage{amsmath, amsfonts, amssymb, amsthm, nicefrac, bbm, bm, stmaryrd}
\newtheorem{thm}{Theorem}[]
\newtheorem{prop}[thm]{Proposition}
\newtheorem{cor}[thm]{Corollary}
\newtheorem*{corbehr}{Corollary 2.2 in \cite{behr:etal:13}}
\newtheorem{lemma}[thm]{Lemma}
\theoremstyle{remark}
\newtheorem*{rmk}{Remark}

\usepackage{algorithm, algpseudocode}

\def\CC{{C\nolinebreak[4]\hspace{-.05em}\raisebox{.4ex}{\tiny\bf ++}}}

\usepackage{soul}

\usepackage{xcolor}
	\definecolor{verde1}{HTML}{00c800}
	\definecolor{verde2}{HTML}{005c00}
	\definecolor{verde3}{HTML}{00ff84}
	\definecolor{blu1}{HTML}{0081ff}
	\definecolor{blu2}{HTML}{0000ff}
	\definecolor{blu3}{HTML}{00eaff}
	\definecolor{rosso1}{HTML}{ff2a00}
	\definecolor{acqua1}{HTML}{009199}
\newcommand{\blue}{\textcolor{black}}

\usepackage{lscape, afterpage}

\usepackage{empheq}

\usepackage{tcolorbox}
\tcbuselibrary{many}
\tcbset{
	enhanced,
	breakable,
	attach boxed title to top left={
		xshift=0.5cm,
		yshift=-\tcboxedtitleheight/2},
	top=4mm,
	beforeafter skip=2\baselineskip,
	colframe = verde2,
	colback = white,
	colbacktitle = green,
	coltitle = black,
	boxrule = 2pt
}

\usepackage{tikz, pgfplots}
\pgfplotsset{compat = 1.14}
\usepgflibrary{shapes.arrows}
\usepgfplotslibrary{statistics}
\usetikzlibrary{arrows,shapes,positioning}

\tikzset{,
	st1/.style = {rectangle, draw=blu2, thick, minimum width=2.75cm, minimum height=1.75cm},
	st2/.style = {rectangle, draw=blu2, thick, minimum width=2.25cm, minimum height=1.75cm},
	st3/.style = {rectangle, draw=blu2, thick, minimum width=3.75cm, minimum height=1.75cm},
	st4/.style = {rectangle, draw=blu2, thick, minimum width=3.25cm, minimum height=1.75cm},
	st5/.style = {ellipse , draw=verde2, thick, minimum height=1cm},
}

\makeatletter
\def\clineThicknessColor#1#2#3{\@ClineThicknessColor#1\@nil{#2}{#3}}
\def\@ClineThicknessColor#1-#2\@nil#3#4{%
	\omit
	\@multicnt#1%
	\advance\@multispan\m@ne
	\ifnum\@multicnt=\@ne\@firstofone{&\omit}\fi
	\@multicnt#2%
	\advance\@multicnt-#1%
	\advance\@multispan\@ne
	{\color{#4}%
		\leaders\hrule\@height#3\hfill}%
	\cr}
\makeatother

\newcommand{\V}{\mathcal{V}}
\newcommand{\E}{\mathcal{E}}
\newcommand{\Vm}{\mathcal{V}^{(m)}}
\newcommand{\Em}{\mathcal{E}^{(m)}}

\newcommand{\bY}{\bm{Y}}
\newcommand{\bZ}{\bm{Z}}
\newcommand{\Ym}{Y_{i_1, \dots, i_m}}
\newcommand{\ym}{y_{i_1, \dots, i_m}}
\newcommand{\Bqm}{B^{(m,n)}_{q_1, \dots, q_m}}
\newcommand{\Bqim}{B^{(m,n)}_{q_{i_1}, \dots, q_{i_m}}}

\newcommand{\PP}{\mathbb{P}}
\newcommand{\Pt}{\mathbb{P}_\theta}
\newcommand{\Et}{\mathbb{E}_\theta}
\newcommand{\Qt}{\mathbb{Q}_{\tau}}
\newcommand{\EQ}{\mathbb{E}_{\Qt}}

\newcommand{\mh}{multiset hypergraphs}
\newcommand{\mmh}{multiple hypergraphs}
\newcommand\rank{\operatorname{rank}}

\DeclareGraphicsExtensions{.pdf, .jpg, .jpeg, .png, .gif,.eps}
\graphicspath{{figs/}{.}}

\title{Model-based clustering in simple hypergraphs through a stochastic blockmodel}

\author[1]{Luca Brusa}
\author[2]{Catherine Matias}

\affil[1]{ Department of Statistics and Quantitative Methods, University of Milano-Bicocca, Via Bicocca degli Arcimboldi 8, 20100 Milano, Italy. Email: \href{mailto:luca.brusa@unimib.it}{\texttt{luca.brusa@unimib.it}}}
\affil[2]{ Sorbonne Université, Université de Paris Cité, Centre National de la Recherche Scientifique, Laboratoire de Probabilités, Statistique et Modélisation, 4 place Jussieu, 75252 PARIS Cedex 05, France. Email: \href{mailto:catherine.matias@cnrs.fr}{\texttt{catherine.matias@cnrs.fr}}}

\date{}%\today}

\begin{document}
	
	\maketitle
	
   \begin{abstract}
We propose a model to address the overlooked problem of node clustering in simple hypergraphs. Simple hypergraphs are suitable when a node may not appear multiple times in the same hyperedge, such as in co-authorship datasets. Our model generalizes the stochastic blockmodel for graphs and assumes the existence of latent node groups and  hyperedges are conditionally independent given these groups. We first establish the generic identifiability of the model parameters. We then develop a variational approximation Expectation-Maximization algorithm for parameter inference and node clustering, and derive a statistical criterion for model selection.

To illustrate the performance of our \texttt{R} package \texttt{HyperSBM}, we compare it with other node clustering methods using synthetic data generated from the model, as well as from a line clustering experiment and a co-authorship dataset. 
\\
        
    \end{abstract}

\section{Introduction}\label{sec:intro}
Over the past two decades, a wide range of models has been developed to capture pairwise interactions represented in graphs. However, modern applications in various fields have highlighted the necessity to consider high-order interactions, which involve groups of three or more nodes. Simple examples include triadic and larger group interactions in social networks \citep[whose importance has been recognized early on, see][]{simm:50}, scientific co-authorship \citep{estr:rodr:06}, interactions among more than two species in ecological systems \citep{muyi:deba:rao:20,sing:baru:21}, or high-order correlations between neurons in brain networks \citep{chela:21}. To formalize these high-order interactions, hypergraphs provide the most general framework. Similar to a graph, a hypergraph consists of a set of nodes and a set of hyperedges, where each hyperedge is a subset of nodes involved in an interaction.

In this context, it is important to distinguish simple hypergraphs from \textit{multiset hypergraphs}, where hyperedges can contain repeated nodes. Multisets are a generalization of sets, allowing elements to appear with varying multiplicities. Recent reviews on high-order interactions can be found in the works of \citet{batt:cenc:iaco:etal:20}, \citet{bick:gross:harr:etal:21}, and \citet{torr:blev:bass:etal:21}.\\

Despite the increasing interest in high-order interactions, the statistical literature on this topic remains limited. Some graph-based statistics, such as centrality or clustering coefficient, have been extended to hypergraphs to aid in understanding the structure and extracting information from the data \citep{estr:rodr:06}. However, these statistics do not fulfill the need for random hypergraph models.

Early analyses of hypergraphs have relied on their embedding into the space of bipartite graphs \citep[see, e.g.,][]{batt:cenc:iaco:etal:20}. Hypergraphs with self-loops and multiple hyperedges (weighted hyperedges with integer-valued weights) are equivalent to bipartite graphs. However, bipartite graph models were not specifically designed for hypergraphs and may introduce artifacts; we refer to Section~\ref{sec:bipartite} in the Supplementary Material for more details.

Generalizing Erdős-Rényi's model of random graphs leads to uniformly random hypergraphs. This model involves drawing uniformly at random from the set of all $m$-uniform hypergraphs (hypergraphs with hyperedges of fixed cardinality $m$) over a set of $n$ nodes. However, similar to Erdős-Rényi's model for graphs, this hypergraph model is too simplistic and homogeneous to be used for statistical analysis of real-world datasets.
In the configuration model for random graphs, the graphs are generated by drawing uniformly at random from the set of all possible graphs over a set of $n$ nodes, while satisfying a given prescribed degree sequence. In the context of hypergraphs, configuration models were proposed in \cite{ghos:zlat:cald:etal:09}, focusing on tripartite and 3-uniform hypergraphs. Later, \cite{chod:20} extended the configuration model to a more general hypergraph setup. In these references, both the node degrees and the hyperedge sizes are kept fixed (a consequence of the fact that they rely on bipartite representations of hypergraphs).
The configuration model is useful for sampling (hyper)graphs with the same degree sequence (and hyperedge sizes) as an observed dataset through shuffling algorithms. Therefore, it is often employed as a null model in statistical analyses. However, sampling exactly (rather than approximately) from this model poses challenges, particularly in the case of hypergraphs. For a comprehensive discussion on this issue, we refer readers to Section 4 in \cite{chod:20}.\\

Another popular approach for extracting information from heterogeneous data is clustering.
In the context of graphs, stochastic blockmodels (SBMs) were introduced in the early eighties \citep{fran:hara:82,holl:lask:lein:83} and have since evolved in various directions. These models assume that nodes are grouped into clusters, and the probabilities of connections between nodes are determined by their cluster memberships. Variants of SBMs have been developed to handle weighted graphs and degree-corrected versions, among others.
\blue{In the context of hypergraphs, \cite{ghos:dukk:17:aos} studied a spectral clustering approach based on a hypergraph Laplacian, and obtained  its weak consistency under a Hypergraph SBM (HSBM) under certain restrictions on the model parameters.  
More recently, \cite{deng_etal_22} established the strong consistency of the basic spectral clustering under the degree-corrected HSBM (DCHSBM)  in the sparse regime where the maximum expected hyperdegree might be of order $\Omega(\log n)$ and $n$ is the number of nodes.} 
By introducing hypergraphons, \cite{bala:21} extended the ideas of hypergraph SBMs to a nonparametric setting. In a parallel vein, \cite{turn:luna:neme:airo:21} proposed a latent space model for hypergraphs, generalizing random geometric graphs to hypergraphs, although it was not specifically designed to capture clustering.
An approach linked to SBMs is presented in \cite{vazq:09}, where nodes belong to latent groups and participate in a hyperedge with a probability that depends on both their group and the specific hyperedge.

Modularity is a widely used criterion for clustering entities in the context of interaction data. It aims to identify specific clusters, known as communities, characterized by high within-group connection probabilities and low between-group connection probabilities \citep{ghos:dukk:14}. However, in the hypergraph context, the definition of modularity is not unique.
In particular, \cite{kami:etal:19} introduced a ``strict'' modularity criterion, where only hyperedges with all their nodes belonging to the same group contribute to an increase in modularity. Their criterion measures the deviation of the number of these homogeneous hyperedges from a new null model called the configuration-like model for hypergraphs, where the average values of the degrees are fixed.
Building upon this, \cite{chod:veld:bens:21} proposed a degree-corrected hypergraph SBM and introduced two new modularity criteria. Similar to \cite{kami:etal:19}, one of these criteria utilizes an ``all-or-nothing'' affinity function that distinguishes whether a given hyperedge is entirely contained within a single cluster or not. In this setup, they established a connection between approximate maximum likelihood estimation and their modularity criterion. This work is reminiscent of the work of \cite{newm:16} in the graph context.
%\red{The symmetric modularity introduced by Chodrow depends on too many parameters (affiliation function) and currently, there is no available implementation of the estimation of this function. Thus, currently, they only offer a clustering approach through the all-or-nothing modularity.}
%
However,  the estimators proposed by \cite{chod:veld:bens:21} do not guarantee maximum likelihood estimation, as the parameter space is constrained by assuming a symmetric affinity function. \blue{We refer to  \cite{Poda_Matias:24} for an empirical comparison of these modularity-based methods.}

It is important to highlight that the developments presented in \cite{kami:etal:19} and \cite{chod:veld:bens:21} are specifically conducted in the context of multiset hypergraphs, where hyperedges can contain repeated nodes with certain multiplicities.
The use of multiset hypergraphs simplifies some of the challenges associated with computing modularity. However, to the best of our knowledge, modularity approaches still lack instantiation in the case of simple hypergraphs where each node can only appear once in a hyperedge.
More specifically, the null model used in hypergraph modularity criteria relies on a model for \mh, similar to how the null model used in classical graph modularity is based on graphs with self-loops. While it is known in the case of graphs that this assumption is inadequate, as it induces a stronger deviation than expected and affects sparse networks as well \citep{Massen_Doye05,Cafieri_2010,Squartini_2011}, the assumption of multisets has not yet been discussed in the context of hypergraph modularity. 

In the context of community detection, random walk approaches have also been utilized for hypergraph clustering \citep{swan:zhan:21}. Additionally, low-rank tensor decompositions have been explored \citep{ke:shi:xia:20}.
The misclassification rate for the community detection problem in hypergraphs and its limits have been analyzed in various contexts \citep[see, for instance,][]{ahn:lee:suh:18,chie:lin:wang:19,cole:zhu:20}.
It is worth mentioning that a recent approach has been proposed to cluster hyperedges instead of nodes \citep{ng:murp:21}. However, our focus in this work is on clustering nodes.

The literature on high-order interactions often discusses simplicial complexes alongside hypergraphs \citep{batt:cenc:iaco:etal:20}. However, the unique characteristic of simplicial complexes, where each subset of an occurring interaction should also occur, places them outside the scope of this introduction, which is specifically focused on hypergraphs.\\

In this article, our focus is on model-based clustering for simple hypergraphs, specifically studying stochastic hypergraph blockmodels. 
We formulate a general stochastic blockmodel for simple hypergraphs, along with various submodels
%, and outline the key differences compared to previous proposals 
(Section~\ref{sec:model}). We provide the first result on the generic identifiability of parameters in a hypergraph stochastic blockmodel (Section~\ref{sec:ident}). Parameter inference and node clustering are performed using a variational Expectation-Maximization (\texttt{VEM}) algorithm (Section~\ref{sec:vem}) that approximates the maximum likelihood estimator. Model selection for the number of groups is based on an integrated classification likelihood (ICL) criterion (Section~\ref{sec:ICL}).
To illustrate the performance of our method, we conduct experiments on synthetic sparse hypergraphs, including a comparison with hypergraph spectral clustering (HSC) and modularity approaches (Section~\ref{sec:simus}). Notably, the line clustering experiment (Section~\ref{sec:line}) highlights the significant differences between our approach and the one proposed by \cite{chod:veld:bens:21}. 
%\red{As a by-product, our synthetic experiments demonstrate that the detectability thresholds for non-uniform sparse hypergraphs, also known as Kesten-Stigum thresholds \citep{ange:etal:15,step:zhu:22}  cannot be deduced from the uniform case (Section~\ref{sec:sim_estim}).} 
We also analyze a co-authorship dataset, presenting conclusions that differ from spectral clustering and bipartite stochastic blockmodels (Section~\ref{sec:data}). \blue{We discuss  (Section~\ref{sec:discuss})  our approach, its advantages, current limitations and possible extensions}.  An \texttt{R} package, \texttt{HyperSBM}, which implements our method in efficient \texttt{C++} code, as well as all associated scripts, are available online (see Section~\ref{sec:available}). This manuscript is accompanied by a Supplementary Material (SM) that contains additional information and experiments, \blue{as well as the proofs of all theoretical results}.

%%%%%%%%%%%%%%%
%%%%%%%%%%%%%%%

%%%%%%%%%%%%%

%%%%%%%%%%%%
%%%%%%%%%%%%
\section{A stochastic blockmodel for hypergraphs}\label{sec:HSBM}
       
\subsection{Model formulation}
\label{sec:model}  
Let $ \mathcal{H}=(\V,\E) $ represent a binary hypergraph, where $ \V = \{ 1, \dots, n \} $ is a set of $ n $ nodes and $ \E $ is the set of hyperedges. In this context, a hyperedge of size $ m \geq 2 $ is defined as a collection of $ m $ distinct nodes from $ \V $. We do not allow for hyperedges to be multisets or self-loops. Let $ M = \underset{e \in \E}{\max} |e| $ denote the largest possible size of hyperedges in $ \mathcal{E} $, with $ M \geq 2 $ (for graphs, $ M = 2 $).
We  define the sets of  \blue{(unordered)} node subsets,  \blue{(ordered)}  node tuples, and hyperedges of size $ m $ as 
\begin{align*}
\Vm &= \big\{ \{ i_1, \dots, i_m \} : i_1, \dots, i_m \in \V \text{ are all distinct} \big\} ,\\
\V^m & = \big\{ ( i_1, \dots, i_m) : i_1, \dots, i_m \in \V \text{ are all distinct} \big\} ,\\
\Em &= \big\{ \{ i_1, \dots, i_m \} \in \Vm : \{ i_1, \dots, i_m \} \in \mathcal{E} \big\}, 
\end{align*}
respectively.
%%%
Obviously $ \E =\bigcup_{m = 2}^{M} \Em \subseteq \bigcup_{m = 2}^{M} \Vm $. 
%%%
For each node subset $ \{ i_1, \dots, i_m \} \in \Vm $, we  define the indicator variable:
\begin{equation*}
\Ym = \mathbbm{1}_{\{ i_1, \dots, i_m \} \in \mathcal{E}} = \begin{cases}
1 \qquad	&\text{if } \{ i_1, \dots, i_m \} \in \mathcal{E},\\
0 				&\text{if } \{ i_1, \dots, i_m \} \notin \mathcal{E}.
\end{cases}
\end{equation*}
We  represent a random hypergraph as $\bY=(\Ym)_{{i_1,\dots,i_m}\in \Vm, 2\le m \le M}$.

Similar to the formulation of the stochastic blockmodel (SBM) for graphs, we assume that the nodes in the hypergraph belong to $ Q $ unobserved groups. We use $ Z_1, \dots, Z_n $ to denote $ n $ independent and identically distributed latent variables, where $ Z_i $ follows a prior distribution $ \pi_q = \mathbb{P}( Z_i = q ) $ for each $ q = 1, \dots, Q $. The values $ \pi_q $ satisfy $ \pi_q \geq 0 $ and $ \sum_{q = 1}^{Q} \pi_q = 1 $. To simplify notation, we sometimes represent $ Z_i $ as a binary vector $ Z_i = (Z_{i1}, \dots, Z_{iQ}) \in \{ 0, 1 \}^Q $, where only one element, $ Z_{iq} $, equals 1. We also define $ \bZ=(Z_1,\dots,Z_n) $.
\blue{
Every $m$-subset of nodes $\{ i_1, \dots, i_m \}$ in $\Vm$ is associated to a latent configuration, namely a set  $\{Z_{i_1},\dots,Z_{i_m}\} = \{ q_1, \dots, q_m \}$ of latent groups to which these nodes belong.	 
%We define  
%\begin{equation*}
%\Qm= \big\{ \{ q_1, \dots, q_m \} : q_1, \dots, q_m \in \{ 1, \dots, Q \} \big\}, 
%\end{equation*}
%as the set of all possible latent configurations for elements in $\Vm$. Note that 
The values of the latent groups within a configuration may be repeated, so that each $\{ q_1, \dots, q_m \}$ is a multiset}.
Now, given the latent variables $ \bZ $, all indicator variables $ Y_{i_1, \dots, i_m} $ are assumed to be independent and to follow a Bernoulli distribution whose  parameter  depends on the latent configuration: 
\begin{equation*}
\Ym | \{ Z_{i_1} = q_1, \dots, Z_{i_m} = q_m \} \sim \mathcal{B}(\Bqm),	\qquad \text{ for any } \{ i_1, \dots, i_m \} \in \Vm. 
\end{equation*}
Here \blue{$ \Bqm =B^{(m,n)}_{q_1,\dots,q_m}$} represents the probability that $ m $ unordered nodes, with latent configuration $\{q_1, \dots, q_m\}$, are connected into a hyperedge. \blue{To simplify notation, we drop the superscript $(m,n)$. However, the model may account for 2 possible sparse settings.  First, as the number of nodes $n$ increases, it is natural to assume that the probability of a hyperedge may decrease; otherwise, we would only observe dense hypergraphs. Second, 
%while it is implicit that $ \Bqm $ depends on $ m $ through $ q_1, \dots, q_m $, 
%we emphasize this dependency as 
it is likely that real data contain fewer hyperedges of larger size $ m $.}
Each $ B$ is a fully symmetric tensor of rank $ m $, namely 
\begin{equation}
\label{eq:sym_tensor}
	\Bqm = B_{q_{\sigma(1)}, \dots, q_{\sigma(m)}}, \ \forall q_1, \dots, q_m \text{ and } \forall \sigma \text{ permutation of } \{1,\dots,m\}.
\end{equation}
%%%%%%
We denote the parameter vector as  $ \theta = (\pi_q, \Bqm)_{q,m,q_1\le \dots \le q_m} $ and the corresponding probability distribution and expectation as  $\Pt,\Et$, respectively. 

\begin{lemma}
\label{lem:nb}
The number of different parameters in each tensor $ B=(\Bqm)_{1\le  q_1\le \dots \le q_m\le Q}$ is  $\binom{Q+m-1}{m}$.
\end{lemma}
As a result, the total number of parameters in our hypergraph stochastic blockmodel (HSBM) is given by:
\begin{equation*}
 (Q-1) + \sum_{m = 2}^{M} \binom{Q+m-1}{m}.
\end{equation*}
As shown in Table~\ref{tab:n_pars}, the number of $\Bqm$ parameters increases  rapidly as the values of $ Q $ and $ m $ grow. 
\blue{Note that the number of parameters (of the order $O(MQ^M+Q)$) remains small compared to the number of observations ($\sum_{m=2}^M\binom n m =O( n^M)$). So we do have enough statistical information to estimate all parameters. Nonetheless,}
to reduce the complexity of the model, we introduce submodels by assuming equality of certain conditional probabilities  $ \Bqm $. 
%We mention that \cite{chod:veld:bens:21} have also defined submodels in the context of degree-corrected HSBM.
%%
In particular, we consider two \emph{affiliation}  submodels given by 
	\begin{equation}
	\label{eq:aff-m}
	\Bqm = \begin{cases}
			\alpha^{(m)}	\quad \text{if } q_1 = \dots = q_m,  \\
			\beta^{(m)}		\quad \text{if there exist at least } q_i \neq q_j \text{ for } i \neq j 
		\end{cases} \tag{\textbf{Aff-m}}
	\end{equation}
and 
	\begin{equation}
	\label{eq:aff}
		\Bqm = \begin{cases}
			\alpha			\quad \text{if } q_1 = \dots = q_m \\
			\beta			\quad \text{if there exist at least } q_i \neq q_j \text{ for } i \neq j 
		\end{cases}  \forall m = 2, \dots, M. \tag{\textbf{Aff}}
	\end{equation}
The number of parameters is dropped to $ (Q-1) + 2(M-1) $  and to $ (Q-1) + 2 $ under Assumptions~\eqref{eq:aff-m} and \eqref{eq:aff}, respectively. 
These submodels align with the concepts discussed in \citet{kami:etal:19} and \citet{chod:veld:bens:21}, where they propose that only hyperedges with nodes belonging to the same group should contribute to the increase in modularity. Additionally, \blue{when $\alpha^{(m)}>\beta^{(m)}$ (resp. $\alpha>\beta$)} these submodels correspond to the scenarios in which \cite{ghos:dukk:14,ghos:dukk:17:aos} obtained their results.

%%%%%%%
% Table for the number of parameters
%%%%%%
%	\bigskip
	\begin{table}[htbp]
		\caption{Number $\binom{Q+m-1}{m}$ of connectivity parameters $\Bqm$ of the full HSBM for given values of $Q$ (number of latent groups) and $m$ (hyperedge size).}
		\centering
		\begin{tabular}{p{1cm}C{1cm}C{1cm}C{1cm}C{1cm}C{1cm}C{1cm}}
			\noalign{\smallskip}\noalign{\hrule height .03cm}\noalign{\smallskip}
					& \multicolumn{6}{c}{$ Q $}						\\
			\cmidrule(r){2-7}
			$ m $	& 2		& 3 	& 4 	& 5		& 6 	& 7 	\\
			\midrule
			3 		& 4		& 10  	& 20  	& 35  	& 56   	& 84	\\
			4 		& 5 	& 15  	& 35  	& 70 	& 126  	& 210	\\
			5 		& 6 	& 21  	& 56 	& 126 	& 252  	& 462	\\
			6 		& 7 	& 28  	& 84 	& 210 	& 462 	& 924	\\
			7 		& 8 	& 36 	& 120 	& 330 	& 792 	& 1716	\\
			\noalign{\smallskip}\noalign{\hrule height .03cm}\noalign{\smallskip}
		\end{tabular}
		\label{tab:n_pars}
	\end{table}
%	\bigskip

\blue{A summary of the manuscript notation is given in Table~\ref{tab:notation}.} 

%%%%%%%%%%%%
% Table of notation
\begin{table}[htbp]
    \caption{\blue{Notation summary.}}
    \centering
    \begin{tabular}{c|p{10cm}}
     $ \mathcal{H}=(\V,\E) $  &  hypergraph with $\V=\{1,\dots,n\}$ set of nodes and $\E$ collection  of (simple) hyperedges\\
     $ M, Q$ & largest hyperedge size and number of clusters\\
     $ \Vm   $ & node subsets (unordered) of size $2\le m\le M$ \\
     $\V^m$ & node tuples (ordered) of size $2\le m\le M$  \\
      $\Em $& hyperedges of size $2\le m\le M$\\
%      $\Qm $& all latent configurations for node subsets of size $2\le m\le M$\\ % REMOVED
      $\bY=(\Ym)_{\{i_1,\dots,i_m\}\in \Vm, 2\le m \le M}$ & observations (presence/absence of a hyperedge at each node subset)\\
      $\bZ=(Z_1,\dots, Z_n)$ & latent configurations (latent clusters), each $Z_i\in \{1,\dots, Q\}$ \\
     $ \pi_q = \mathbb{P}( Z_i = q ) \in [0,1]$ & clusters proportions, such that $\sum_{q=1}^Q \pi_q= 1$\\
      $\Bqm\in [0,1]$ &  probability of a hyperedge at  a size-$m$ node subset with latent configuration $\{q_1,\dots,q_m\},$ for $1\le q_1\le \dots\le q_m\le Q$\\
       $\theta = (\pi_q, \Bqm)_{ q, m, q_1\le \dots \le q_m}$ & model parameter \\
      $\alpha^{(m)}, \beta^{(m)}$ (resp. $\alpha, \beta$) & within-clusters and between-clusters probabilities in the affiliation sub-model~\eqref{eq:aff-m} (resp. \eqref{eq:aff}) \\
      $\Qt(\cdot)$ & variational distribution on the latent configurations $\bZ$\\
      $\tau_{iq}\in [0,1]$ & variational probability that node $i$ belongs to cluster $q$, such that $\sum_{q=1}^Q \tau_{iq}=1$ for all $i\in \{1,\dots,n\}$\\
      $f (y, b) $ & Bernoulli density at $y$ with parameter $b$\\
      $\mathcal{J}(\theta, \tau)$ & evidence lower bound (ELBO)\\
      $\mathcal{H}(\cdot), \text{KL}(\cdot || \cdot)$ & entropy and Kullback-Leibler divergence
    \end{tabular}
    \label{tab:notation}
\end{table}

%%%%%%%%%%%%%%%%
%%%%%%%%%%%%%%%%
\subsection{Parameter identifiability}\label{sec:ident}
We first establish the generic identifiability of the parameter in a HSBM that is restricted to simple $m$-uniform hypergraphs for any $m\geq 2$. In a parametric context, generic identifiability implies \blue{that the distribution $\Pt$ of a hypergraph over a set of $n$ nodes uniquely defines the parameter $\theta$, except possibly for some parameters in a subset of dimension strictly smaller than the full parameter space. In other words,  if we randomly select a parameter $\theta \in \Theta$ according to the Lebesgue measure, the distribution $\Pt$ uniquely characterizes the parameter $\theta$, for a large enough number of nodes $n$.}
%that every parameter $\theta$, except possibly for some parameters in a subset of dimension strictly smaller than the full parameter space, uniquely determines the distribution  $\Pt$. In other words,  if we randomly select a parameter $\theta \in \Theta$ according to the Lebesgue measure, it almost surely uniquely defines $\Pt$.
Identifiability is established up to label switching on the node groups, as is common in discrete latent variable models. 
For the case of $m=2$, the identifiability result corresponds to Theorem 2 in \cite{allm:etal:11}. Our proof follows similar principles, building upon a key result by \cite{krus:77}. In our case, we crucially additionally rely on a sufficient condition for a sequence of nonnegative integers to represent the degree sequence of a simple $m$-uniform hypergraph, as established by \cite{behr:etal:13}.

\begin{thm}\label{thm:ident}
  For any $m\ge 2$ and any integer $Q$, 
  the parameter $\theta=(\pi_q, \Bqm)_{1\le q\le Q, 1\le q_1\le \dots \le q_m \le Q}$ of the HSBM restricted to
  $m$-uniform simple  hypergraphs over $n$ nodes, is  generically identifiable,  up to label switching on the node
  groups, \blue{as soon as $n \ge Q^2 \Big(m! Qm +m -1\Big)^{2/(m-1)}$.}
  %for large enough $n$ (depending only on $m,Q$).
 Moreover, the result remains valid when the group proportions $\pi_q$ are fixed.
\end{thm}

\blue{This result does not provide specific insights into the identifiability in 
%Similarly, this result says nothing about 
the affiliation cases \eqref{eq:aff-m} and \eqref{eq:aff}. Indeed, it does not explicitly characterize the subspace of the parameter space where identifiability may not hold, although we know that its dimension is smaller than that of the full parameter space (and that the possible restrictions apply only on the part of the parameter space concerning the probabilities of connection $\Bqm$).}
%which correspond to further restrictions of the parameter space to lower-dimensional subspaces.}

%
The result we have established for $m$-uniform hypergraphs also implies a similar result for non-uniform simple hypergraphs, as shown in the following corollary.

\begin{cor}\label{cor:ident}
  For any integer $Q$, 
  the parameter $\theta=(\pi_q, \Bqm)_{1\le q\le Q, 1\le q_1\le \dots \le q_m \le Q, 2\le m\le M}$ of the HSBM for simple hypergraphs over $n$ nodes is
  generically identifiable,  up to label switching on the node groups,  \blue{as soon as $n \ge Q^2 \Big(M! QM +M -1\Big)^{2/(M-1)}$.}
  %for large enough $n$ (depending only on $M,Q$).
  \end{cor}

Our proof of Corollary~\ref{cor:ident} relies on the assumption that all the $\pi_q$'s are distinct, which is a generic condition. This condition is not explicitly stated in the corollary, but it is required for the proof to hold. Consequently, the result of generic identifiability does not bring any insight in cases where the group proportions are equal, as it is not sufficient to identify the parameters separately for each value of $m$.

Additional technical work is thus needed to establish whether a HSBM with equal group proportions, or whether the affiliation submodels have identifiable parameters. 

\blue{As a final note, we mention that there is no direct link between parameter identifiability and detectability thresholds for clusters recovery \citep{dumitriu2022,step:zhu:22}.%discussed in Section~\ref{sec:sim_estim}. 
While clusters recovery is an asymptotic result with guarantees when the sample size increases,
 parameter identifiability is a theoretical result stating that the distribution of the observations (for a minimal sample size) fully characterizes the parameter. It is theoretical in the sense that it does not deal with inference, though the property has consequences on inference results. 
 Parameter identifiability is a basic requirement for  consistency results of maximum likelihood estimators to hold in parametric settings and it is also required for proofs of clusters exact recovery.}

%%%%%%%%%%%%%%%%
%%%%%%%%%%%%%%%%
\subsection{Parameter estimation via variational Expectation-Maximization}
\label{sec:vem}	
The likelihood of the model is given \blue{as a marginal distribution}  
\begin{align}
\label{eq:lik}
\Pt(\bY) &= \sum_{q_1=1}^Q \dots \sum_{ q_n =1}^Q \Pt(\bY, \bZ=(q_1,\dots,q_n))\nonumber \\
      &= \sum_{q_1=1}^Q \dots \sum_{ q_n =1}^Q
   \left(\prod_{i=1}^n   \Pt (Z_i=q_i) \right)\prod_{m = 2}^{M} \prod_{\{i_1, \dots, i_m\} \in \Vm} \Pt(\Ym | Z_{i_1} = q_{i_1}, \dots, Z_{i_m} = q_{i_m}) \nonumber \\
		&= \sum_{q_1=1}^Q \dots \sum_{ q_n =1}^Q  \left( \prod_{i=1}^n \pi_{q_i}  \right)\prod_{m = 2}^{M} \prod_{\{i_1, \dots, i_m\} \in \Vm } (\Bqim)^{\Ym} (1-\Bqim)^{1-\Ym}.
\end{align}
The computation of the model likelihood is generally intractable. Equation~\eqref{eq:lik} involves a summation over all possible $Q^n$ different latent configurations of the nodes, which becomes computationally prohibitive when $n$ and $Q$ are large.
In the context of latent variable models, the Expectation-Maximization (EM) algorithm \citep{demp:lair:rubi:77} is commonly used to address this issue. However, the \texttt{EM} algorithm cannot be directly applied to SBMs. This is because the \texttt{E}-step, which involves computing the conditional posterior distribution of the latent variables $P(\bZ | \bY)$, is itself intractable in SBMs \citep[see, e.g.,][]{mati:robi:14}.
One possible solution is to employ variational approximations of the \texttt{EM} algorithm, known as Variational \texttt{EM}  \citep[\texttt{VEM,}][]{jord:etal:99}. \blue{Below, we recall the classical approach for the \texttt{VEM} algorithm.}

We \blue{denote the density of the Bernoulli distribution with parameter $b$} as  
\begin{equation}
    \label{eq:density}
\blue{\forall y \in \{0,1\}, \quad} f(y,b):=y\log b +(1-y)\log (1-b).
\end{equation}
Then, the complete data log-likelihood is 
\begin{align}
\label{eq:complete_loglik}
\ell_n^c(\theta) &= \log\Pt(\bY,\bZ) = \log\Pt(\bZ) + \log\Pt(\bY | \bZ)\\
			&= \sum_{q=1}^{Q} \sum_{i=1}^{n} Z_{iq} \log\pi_q  + \sum_{m = 2}^{M} \sum_{q_1=1}^Q \dots \sum_{q_m=1}^Q\sum_{\{i_1,\dots, i_m\}\in \Vm}  Z_{i_1q_1} \cdots Z_{i_mq_m} f(Y_{i_1 \dots i_m}, \Bqm) \nonumber \\
            &= \sum_{q=1}^{Q} \sum_{i=1}^{n} Z_{iq} \log\pi_q 
			 + \sum_{m = 2}^{M} \sum_{q_1\le q_2\le \dots \le q_m}\sum_{(i_1,\dots,i_m)\in \V^m}  Z_{i_1q_1} \cdots Z_{i_mq_m}  f(Y_{i_1 \dots i_m}, \Bqm). \nonumber
\end{align}
Note that the final equality ensures that each parameter value appears only once.
The key principle underlying the variational method is to adopt the same iterative two-step structure as the \texttt{EM} algorithm but replace the intractable posterior distribution $\Pt(\bZ | \bY)$ with the best approximation, in terms of Kullback-Leibler divergence, from a class of simpler distributions, often factorized.
We introduce a class of  probability distributions $\Qt$ over $\bZ=(Z_1, \dots, Z_n)$ \blue{that factorize over the set of nodes, thus} given by
\begin{equation*}
\Qt(\bZ) = \prod_{i=1}^{n} \Qt(Z_{i})
    = \prod_{i=1}^{n} \prod_{q=1}^Q \tau_{iq}^{Z_{iq}},
\end{equation*}
where the variational parameter $\tau_{iq} = \Qt(Z_i=q) \in [0,1]$ satisfies $\sum_{q=1}^{Q} \tau_{iq} = 1$ for any $i = 1, \dots, n$.
%and $q = 1, \dots, Q$.
The expectation under distribution $\Qt$ is denoted as $\EQ$, and $\mathcal{H}(\Qt)$ represents the entropy of $\Qt$. 
Now we  define the evidence lower bound (ELBO): 
\begin{align}\label{eq:elbo_def}
\mathcal{J}(\theta, \tau)
&= \EQ [\log\Pt (\bY,\bZ)] + \mathcal{H}(\Qt) \\
&= \EQ [\log\Pt (\bY,\bZ)] - \EQ[\log \Qt (\bZ)] \nonumber\\
&= \sum_{q=1}^{Q}\sum_{i=1}^{n}\tau_{iq}\log\frac{\pi_q}{\tau_{iq}}  + \sum_{m=2}^{M}
\sum_{q_1\le q_2\le \dots \le q_m} \sum_{(i_1,\dots,i_m)\in \V^m} 
\tau_{i_1q_1} \cdots \tau_{i_mq_m} f(Y_{i_1 \dots i_m}, \Bqm)\nonumber.
\end{align}
It can be observed that 	 $ \mathcal{J}(\theta, {\tau}) $ satisfies the following relation:  
\begin{equation}
\label{eq:elbo_prop}
\mathcal{J}({\theta}, {\tau}) = \log\Pt(\bY) - \text{KL}(\Qt(\bZ) || \Pt(\bZ | \bY)),
\end{equation}
where $\text{KL}(\cdot || \cdot)$ denotes the Kullback-Leibler divergence. \blue{Equation~\eqref{eq:elbo_prop} is at the core of the \texttt{EM} algorithm and its variational approximation. In the classical \texttt{EM} approach, at the $t$-th iteration of the algorithm, the variational distribution $\Qt$ is chosen as the distribution $\mathbb{P}_{\theta^{(t)}}(\bZ | \bY)$ of the latent variables given the observations at the current parameter value $\theta^{(t)}$. This cancels the Kullback-Leibler term and the ELBO equals the log-likelihood. When the distribution $\Pt(\bZ | \bY)$ is not factorized, such a choice would prevent from an efficient computation of the expectation $\EQ [\log\Pt (\bY,\bZ)]$ of the complete log-likelihood under $\Qt$ appearing in Equation~\eqref{eq:elbo_def}. Thus, the variational approximation searches for the best approximation of the true $\Pt (\bZ | \bY)$ in a class of simplified (in general, factorized) variational distributions. As a consequence, the Kullback-Leibler divergence term in~\eqref{eq:elbo_prop} is non null in general and} the ELBO $\mathcal{J}$ serves as a lower bound for the model log-likelihood $\log\Pt(\bY)$.  
The \texttt{VEM} algorithm iterates between the following two steps until a suitable convergence criterion is met:
\begin{itemize}
\item {\bf \texttt{VE}-Step} maximizes $ \mathcal{J}({\theta}, {\tau}) $ with respect to $ {\tau} $
\begin{equation}
    \label{eq:ve_step}
\widehat{{\tau}}^{(t)} = \underset{{\tau}}{\arg\max}\ \mathcal{J}({\theta}^{(t-1)}, {\tau}), \quad \text{s.t. } \textstyle\sum_{q=1}^{Q} \widehat{\tau}^{(t)}_{iq} = 1 \qquad \forall i = 1, \dots, n . 
\end{equation}
This step involves finding the best approximation of the conditional distribution $\Pt(\bZ|\bY)$ by minimizing the Kullback-Leibler divergence term in \eqref{eq:elbo_prop}.
\item {\bf \texttt{M}-Step} maximizes $ \mathcal{J}({\theta}, {\tau}) $ with respect to $ {\theta} $
\begin{equation}\label{eq:m_step}
\widehat{{\theta}}^{(t)} = \underset{{\theta}}{\arg\max}\ \mathcal{J}({\theta}, {\tau}^{(t-1)}),\quad \text{s.t. } \textstyle\sum_{q=1}^{Q} \widehat{\pi}^{(t)}_q = 1 .
\end{equation}
This step updates the values of the model parameters $\pi_q$ and $\Bqm$.
\end{itemize}
	
In the following we provide the solutions of the two maximization problems in Equations~\eqref{eq:ve_step} and \eqref{eq:m_step}.
	
\begin{prop}[\texttt{VE}-Step]
\label{prop:VEstep}
Given the current model parameters $ \theta= (\pi_q , \Bqm )_{q,m,q_1\le \dots \le q_m} $ at any iteration of the \texttt{VEM} algorithm, the corresponding optimal values of the variational parameters $ (\widehat{\tau}_{iq})_{i,q} $ defined in Equation~\eqref{eq:ve_step}  satisfy the  fixed point equation: 
\begin{equation}
\label{eq:fixed_point}
\log \widehat{\tau}_{iq} =\log \pi_q +\sum_{m=1}^{M-1}
\sum_{1\le q_1\le  \dots \le q_m\le Q} \sum_{\substack{(i_1,\dots,i_m)\in \V^m \\ \text{s.t.} \{i,i_1,\dots,i_m\} \in \V^{(m+1)}}}
\widehat{\tau}_{i_1q_1}\cdots\widehat{\tau}_{i_mq_m} f(Y_{ii_1 \dots i_m}, B^{(m+1,n)}_{qq_1 \dots q_m})
+c_i, 
\end{equation}
for any $1\le i\le n$ and $1\le q \le Q$ and 
where $c_i$ are normalising constants such that $\sum_q \widehat{\tau}_{iq}=1 $.
\end{prop}

\blue{Equation~\eqref{eq:fixed_point} relates the variational probability $\widehat{\tau}_{iq}$ that a node $i$ belongs to a cluster $q$ to the other variational parameters (as well as the observations and current parameter value $\theta$). The sum starts at $m=1$ and deals with $(m+1)$-tuples of nodes $\{i, ,i_1,\dots,i_m\}$ that contain node $i$ and whose latent configuration is given by some multiset $\{q, q_1,\dots, q_m\}$.}

\begin{rmk}
From Proposition~\ref{prop:VEstep}, the $\tau_i$'s are obtained using a fixed point algorithm. Although in all the situations we experienced, the algorithm converged in a reasonable number of iterations, we have no guarantee about existence nor uniqueness of a solution to \eqref{eq:fixed_point}.
\end{rmk}

%%%%	
\begin{prop}[\texttt{M}-Step]
\label{prop:Mstep}
Given the variational parameters $ (\tau_{iq})_{i,q} $ at any iteration of the \texttt{VEM} algorithm, the corresponding optimal values of the model parameters $ (\widehat{\pi}_q , \widehat{B}_{q_1 \dots q_m})_{q,m,q_1 \le \dots\le  q_m} $ defined in Equation~\eqref{eq:m_step} are given by 
\begin{equation*}
\widehat{\pi}_q = \frac{1}{n} \sum_{i=1}^{n}\tau_{iq} \qquad \text{ and } \qquad
\widehat{B}_{q_1 \dots q_m} = \frac{\sum_{(i_1,\dots,i_m)\in \V^m} \tau_{i_1q_1} \dots \tau_{i_mq_m} Y_{i_1 \dots i_m}}{\sum_{(i_1,\dots,i_m)\in \V^m}  \tau_{i_1q_1} \dots \tau_{i_mq_m}}.
\end{equation*}
\end{prop}
%%%%%

We now express the solutions of the \texttt{M}-Step under the submodels given by~\eqref{eq:aff-m} and \eqref{eq:aff}. Note that the \texttt{VE}-Step is unchanged under %LB this setting.
these settings.

\begin{prop}[\texttt{M}-Step, affiliation setups]
\label{prop:Mstep-aff}
In the particular %LB affiliations
affiliation submodels given by~\eqref{eq:aff-m} and \eqref{eq:aff} respectively,  given variational parameters $ (\tau_{iq})_{i,q} $, at any iteration of the \texttt{VEM} algorithm, the corresponding optimal values of $(\widehat {\alpha}^{(m)} , \widehat {\beta}^{(m)})_{m}$ and $\widehat {\alpha} , \widehat {\beta}$ maximising $\mathcal{J}$ as in Equation~\eqref{eq:m_step} are  given by 
\begin{itemize}
    \item Under Assumption~\eqref{eq:aff-m}, 
\begin{align*}
\widehat {\alpha}^{(m)}  &= \frac{\sum_{q=1}^Q\sum_{i_1<\dots<i_m}  \tau_{i_1q} \dots \tau_{i_mq} Y_{i_1 \dots i_m}}{\sum_{q=1}^Q\sum_{i_1<\dots<i_m } \tau_{i_1q} \dots \tau_{i_mq}} , \\
\widehat {\beta}^{(m)} &= \frac{\sum_{\substack{q_1\le \dots \le q_m \\ |\{q_1,\dots,q_m\}|\ge 2}}\sum_{(i_1,\dots,i_m)\in \V^m} \tau_{i_1q_1} \dots \tau_{i_mq_m} Y_{i_1 \dots i_m}}{\sum_{\substack{q_1\le \dots \le q_m \\ |\{q_1,\dots,q_m\}|\ge 2}}\sum_{(i_1,\dots,i_m)\in \V^m} \tau_{i_1q_1} \dots \tau_{i_mq_m}} .
\end{align*}
%%%
\item Under Assumption~\eqref{eq:aff}, 
\begin{align*}
\widehat {\alpha}  &= \frac{\sum_{m=2}^M\sum_{q=1}^Q\sum_{i_1<\dots<i_m} \tau_{i_1q} \dots \tau_{i_m q} Y_{i_1 \dots i_m}} {\sum_{m=2}^M\sum_{q=1}^Q\sum_{i_1<\dots<i_m} \tau_{i_1q} \dots \tau_{i_m q}} , \\
\widehat {\beta} &= \frac{\sum_{m=2}^M\sum_{\substack{q_1\le \dots \le q_m\\ |\{q_1,\dots,q_m\}|\ge 2}}\sum_{(i_1,\dots,i_m)\in \V^m} \tau_{i_1q_1} \dots \tau_{i_m q_m} Y_{i_1 \dots i_m}}{\sum_{m=2}^M\sum_{\substack{q_1\le \dots \le q_m\\ |\{q_1,\dots,q_m\}|\ge 2}}\sum_{(i_1,\dots,i_m)\in \V^m} \tau_{i_1q_1} \dots \tau_{i_m q_m}} .
\end{align*}
\end{itemize}
\end{prop}

\paragraph{Algorithm initialization.}
We choose to begin the algorithm with its \texttt{M}-step, which %LB provides
requires an initial value for $\bm{\tau}$. This allows us to leverage smart initialization strategies based on a preliminary clustering of the nodes. Specifically, we employ three different initialization strategies and select the best result that maximizes the ELBO criterion $\mathcal{J}$:

    \textit{Random initialization}: This naive method involves drawing each $(\tau_{iq})_{1 \leq q \leq Q}$ uniformly from $(0, 1)$ for every node $i$ and normalizing the vector $\tau_i$.

    \textit{``Soft'' spectral clustering}: We utilize Algorithm 1 from \cite{ghos:dukk:17:aos} combined with soft \texttt{$k$-means}. In this approach, we compute a hypergraph Laplacian and construct the column matrix $X$ consisting of its leading $Q$ orthonormal eigenvectors. The rows of $X$ are then normalized to have unit norm  \citep[following steps 1 to 3 in Algorithm 1 from][]{ghos:dukk:17:aos}. We subsequently perform a soft \texttt{$k$-means} algorithm on the rows of $X$ to obtain $\tau_{iq}$, which represents the posterior probability of node $i$ belonging to cluster $q$.

    \textit{Graph-component absolute spectral clustering}: This strategy focuses on edges in the hypergraph ($m=2$) and the corresponding adjacency matrix. We apply the absolute spectral clustering method \citep{rohe:chat:yu:11} to this adjacency matrix. \blue{The absolute spectral clustering method introduces a graph Laplacian with both positive and negative eigenvalues and focuses on the ones with largest magnitude, thus capturing both communities and dis-assortative structures.} It should be noted that this initialization only uses information from hyperedges of size $m=2$, excluding hyperedges with larger sizes. However, absolute spectral clustering is considered superior to spectral clustering as it captures disassortative groups. 
    
\blue{In Section~\ref{app:synth_plus} from the SM, we include a comparison of different initialization strategies. In general,  there won't be an initialization strategy that is always superior, so we recommend always using different strategies and selecting the best criteria. }

\paragraph{Fixed point.}
The \texttt{VE}-Step is achieved through a fixed-point algorithm. In practice, at iteration $t$ of the \texttt{VEM} algorithm, starting from the previous values of the variational and model parameters $\tau_{iq}^{(t-1)}$ and $\theta^{(t-1)}$ respectively, we iterate over some index $u$ to compute the values of $\tau_{iq}^{(t,u)}$ according to Equation \eqref{eq:fixed_point}. This generates a sequence of values $\tau_{iq}^{(t,u)}$. We terminate these iterations either when we reach the maximum number of fixed-point iterations ($u > U_{\max}$) or when the variational parameters have converged ($\underset{iq}{\max} |\tau_{iq}^{(t,u-1)} - \tau_{iq}^{(t,u)}| \leq \varepsilon$), where $\varepsilon$ is a small tolerance threshold.

\paragraph{Stopping criteria.}
The iterations of the \texttt{VEM} algorithm should be terminated when the ELBO $\mathcal{J}$ and the sequence of model parameter vectors $\theta^{(t)}=(\theta^{(t)}_s)_s$ have converged. However, in practice, we have observed that sometimes the algorithm stops prematurely when the \texttt{VE}-Step still requires a few iterations to reach a fixed point. In such cases, continuing with the \texttt{VEM} iterations often leads to higher values of the ELBO function and better parameter estimates. To address this, we enforce the condition that the fixed point in the \texttt{VE}-Step is reached in its first iteration. This reduces the chance of converging to local maxima of $\mathcal{J}$. If these convergence conditions are not met, we stop the algorithm if the maximum number of iterations has been reached.
%%%%%%
    To summarize, we stop the algorithm whenever: 
    \begin{align*}
    & \left\{ %\min \left( 
        \frac{|\mathcal{J}(\theta^{(t-1)}) - \mathcal{J}(\theta^{(t)})|}{|\mathcal{J}(\theta^{(t)})|} \leq \varepsilon %;
        \quad \text{ and } \quad
        \underset{s}{\max} |\theta_s^{(t-1)} - \theta_s^{(t)}| %\right) 
        \leq \varepsilon 
        \quad \text{ and } \quad \underset{iq}{\max} |\tau_{iq}^{(t,0)} - \tau_{iq}^{(t,1)}| \leq \varepsilon \right\} \\
    & \text{ or } \quad \{ t > T_{\max} \}. 
    \end{align*}

Section~\ref{app:details} in SM contains additional details about the algorithm's implementation.

\paragraph{Algorithm complexity and choice of $M$.}
The complexity of our algorithm is of the order 
 $O(nQ^M\binom{n}{M})$,  which can be quite prohibitive for large datasets, especially when $M$ becomes large. 
It is important to emphasize that when analyzing a dataset, the value of $M$ is not necessarily the maximum observed size of the hyperedges, but rather a modeling choice. Indeed, while an occurring hyperedge $\Ym$ with node clusters $\{q_1,\dots, q_m\}$ contributes $\log \Bqm$ to the likelihood, a non occurring one contributes $\log(1-\Bqm)$ and the statistical information that they bring to the parameter is the same (see Equations~\eqref{eq:density} and~\eqref{eq:complete_loglik}). 
Now let's consider for e.g. a co-authorship dataset where we observe  $n$ authors and at most $3$ co-authors per paper. The absence of hyperedges of size $4$ provides as much information for a HSBM as if all possible size-$4$ hyperedges were present. Similarly, the information contained in a dataset with all but 5 possible size-$4$ hyperedges present is the same as the information contained in a dataset with only 5 occurring size-$4$ hyperedges. In other words, $0$ and $1$ values play a symmetric  role. 

As a consequence, the choice of $M$ is left to the discretion of the statistician, depending on the characteristics of the dataset and the available computational resources. In particular, if there are hyperedges with very large sizes $M$, the statistician may decide not to consider them, just as it is justified not to take into account the absence of hyperedges of size $M+1$, where $M$ is the largest observed size. It is important to note that choosing $M>2$ already represents an improvement in terms of considering more information compared to a graph analysis of the data.

 Therefore, for large datasets, we recommend limiting the analysis to smaller values of $M$, such as $M=3$ or $M=4$, to reduce computational burden and improve efficiency.

\subsection{Model selection}
\label{sec:ICL}
While \cite{ghos:dukk:17:aos} propose a method for selecting the number of groups based on the spectral gap, our approach relies on a statistical framework to construct a model selection criterion.

After obtaining the estimated parameters $\hat{\theta}$ and $(\hat{\tau}_i)_i$ from the \texttt{VEM} algorithm, we assign each node $i$ to its estimated group $\hat{Z}_i = \arg \max_q \hat{\tau}_{iq}$. We then define the integrated classification likelihood  \citep[ICL,][]{bier:cele:gova:00} for the full model and the submodels \eqref{eq:aff-m} and \eqref{eq:aff} as follows:

\begin{align*}
\text{ICL}_{\text{full}}(q) &= \log \mathbb{P}_{\hat{\theta}}(\bY, \hat{\bZ}) - \frac{1}{2}(q-1)\log n - \frac{1}{2} \sum_{m=2}^M \binom{q+m-1}{m} \log \binom{n}{m}, \\
\text{ICL}_{\text{aff-m}}(q) &= \log \mathbb{P}_{\hat{\theta}}(\bY, \hat{\bZ}) - \frac{1}{2}(q-1)\log n - (M-1) \sum_{m=2}^M \log \binom{n}{m}, \\
\text{ICL}_{\text{aff}}(q) &= \log \mathbb{P}_{\hat{\theta}}(\bY, \hat{\bZ}) - \frac{1}{2}(q-1)\log n - \sum_{m=2}^M \log \binom{n}{m}.
\end{align*}
\blue{These criteria are constructed as the complete log-likelihood (computed at the estimated parameter value and clusters), penalized by a BIC-like term that accounts for the number of parameters and the corresponding ``effective'' sample size ($n$ for the parameters related to the nodes and $\binom{n}{m}$ for the size-$m$ interaction parameters $\Bqm$). ICL criteria have been widely used in the context of SBMs. Their theoretical properties have never been established, though they exhibit very good empirical results on synthetic SBMs datasets \citep[e.g.][]{Daudin_etal08}. Recently, \cite{Cerqueira_Leonardi_20} obtained a first consistency result for a related criterion in the graph SBM, relying on a penalized version of the exact ICL \citep{Come_Latouche}, also known in the information theory literature as Krichevsky-Trofimov (KT) estimator. While the literature of order estimation focuses on minimal penalties as these will lead to minimum underestimation probability \citep[see for e.g.][]{vanHandel}, 
these KT penalties are generally heavier than what is thought to be sufficient to consistently estimate the number of groups.} 
%%%%%%%%%%%
We determine the number of groups $\hat{q}$ as the value that maximizes the corresponding ICL criterion: $\hat{q} = \arg \max_q \text{ICL}(q)$. %This allows us to select the optimal number of groups based on the statistical properties of the data and the model.

\section{Synthetic experiments}\label{sec:simus}
\subsection{Synthetic data}

We conducted a simulation study to evaluate the performance of the \texttt{HyperSBM} package. 
%Here, we describe the simulation setup and summarize the key results. 
We generated hypergraphs under the HSBM  with two or three latent groups ($Q = 2$ or $Q = 3$). The group proportions were non-uniform, with $\pi = (0.6, 0.4)$ for $Q = 2$ and $\pi = (0.4, 0.3, 0.3)$ for $Q = 3$. We set the largest hyperedge size $M$ to 3, and we considered different numbers of nodes, $n \in \{50, 100, 150, 200\}$.

To simplify the latent structure, we assumed the \eqref{eq:aff-m} submodel, and we parametrized the model through the ratios $\rho^{(m)}$ of within-group size-$m$ hyperedges over between-groups size-$m$ hyperedges (assumed constant with $n$, see Section~\ref{sec_SM:synth_pram} in SM for details). 
We analyzed two different scenarios:  
\begin{enumerate}
    \item[A.]  Communities: in this scenario, we focus on community detection and consider the case of  more within-group than between-groups size-$m$ hyperedges $\rho^{(m)} > 1$ for $m=2,3$. 
    \item[B.] Disassortative: in this scenario, we focus on disassortative behaviour and consider the case of less  within-group than  between-groups size-$m$ hyperedges  $\rho^{(m)} < 1$ for $m=2,3$. 
\end{enumerate}
Each setting is  a combination of a scenario (X=A,B) and number of groups ($Q=2,3$) and is denoted X$Q$. In each setting, values of \blue{$\alpha^{(m)}=\alpha^{(m,n)}$ and $\beta^{(m)}=\alpha^{(m,n)}$ (we here emphasize the dependence on the number of nodes $n$)} decrease with increasing $n$ so as to maintaining constant the quantities $n\alpha^{(2,n)}$ and $n\beta^{(2,n)}$ as well as $n^2\alpha^{(3,n)}$ and $n^2\beta^{(3,n)}$. This implies that the number of size-$m$ hyperedges (both within and between groups) grows linearly with $n$. 
We have explored a total of 5 different settings, denoted by A2, A3, B2, B3  and A3' and we present below the most striking results. In the case of scenarios A (communities) with $Q=3$ groups, we pushed the limits and explore two different settings (namely settings A3 and A3'), with setting A3 being \emph{highly sparse, i.e.} sparser than the already sparse setting A3'.  
Details of the parametrization, specific parameter values and number of hyperedges are fully given in Section~\ref{sec_SM:synth_pram} in SM, while Section~\ref{app:synth_plus} in SM  contains additional results. 

For each setting and each value of $n$, we randomly draw 50 different hypergraphs. We estimate the parameters using the full HSBM formulation with our \texttt{VEM} algorithm, relying on soft spectral clustering (for Scenario A) and graph-component absolute spectral clustering (for Scenario B) initializations (see paragraph ``Algorithm initialization''  above).
%and using the true value for $Q$.

\subsection{Clustering and estimation under HSBM with a fixed number of groups}
\label{sec:sim_estim}
In this part we focus on clustering and parameter estimation with a known number of groups. 
 The performance of \texttt{HyperSBM} is evaluated based on its ability to accurately recover the true clustering and estimate the original parameters. We also  compare our results with hypergraph spectral clustering, relying on Algorithm~1 from~\cite{ghos:dukk:17:aos}, denoted HSC below.

\paragraph*{Clustering results.}
The performance of correct classification is evaluated using the Adjusted Rand Index  \citep[ARI,][]{hube:arab:85}. The ARI measures the similarity between the true node clustering and the estimated clustering. It is upper bounded by 1, where a value of 1 indicates perfect agreement between the clusterings, and negative values indicate less agreement than expected by chance.

\vspace{.5cm}
\begin{figure}[htbp]
    \centering
    \includegraphics[width=\textwidth]{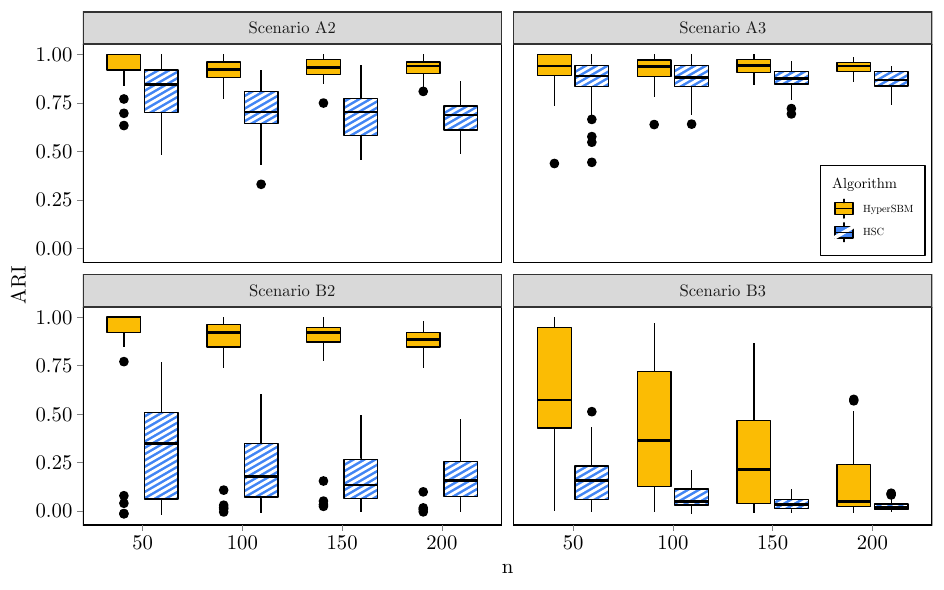}
    \caption{
    Boxplots of Adjusted Rand Indexes for different settings X$Q$ (where X=A, B is the scenario and $Q=  2,3$ is the number of groups),   number of nodes $n$ (along $x$-axis) and 2 methods: our \texttt{HyperSBM} (left boxplot) and HSC (right boxplot). First row (resp. first column) shows scenario A with communities (resp. $Q=2$) while second row (resp. second column) shows scenario B with disassortative behaviour (resp. $Q=3$).}
    \label{fig:ari_res}
\end{figure}
\vspace{.5cm}

Figure~\ref{fig:ari_res} displays the boxplot values of the ARI for settings A2 to B3. 
%computed from 50 simulated datasets. 
It is evident that our \texttt{HyperSBM} consistently outperforms HSC, obtaining higher ARI values overall and significantly lower variances in most cases, except for setting B3, where \texttt{HyperSBM} exhibits a larger variance but still yields substantially better results compared to HSC.
We also observe that increasing the number of nodes $n$ does not appear to significantly enhance the clustering results of \texttt{HyperSBM}. This behavior could be attributed to  our simulation setting, where the numbers of size-$m$ hyperedges ($m=2,3$) are kept linearly increasing with $n$. However, it is worth noting that the variances of the ARI obtained by \texttt{HyperSBM} tend to decrease with an increasing number of nodes $n$. One final comment pertains to the relatively poor clustering performance obtained by both methods in setting B3: this setting appears to be particularly challenging. 
%%% 

%\paragraph*{Detectability thresholds.}
%\red{DROP THIS PART.}
%A threshold for community detection in $m$-uniform sparse hypergraphs is conjectured in \cite{ange:etal:15} and \cite{step:zhu:22} established it is efficiently  achieved by a spectral algorithm based on the non-backtracking operator. We also mention the work by \cite{Zhang_Tan_23} in a different regime. \blue{For non-uniform hypergraphs, fewer results exist and \cite{dumitriu2022}  conjecture a threshold without establishing its validity. Nonetheless, \cite{step:zhu:22} consider that their method should in principle reach that threshold.} 
%In the specific case  of a symmetric \blue{non-uniform HSBM} (which corresponds to \eqref{eq:aff-m} submodel) with equal group proportions \blue{$\pi_q=1/Q$, these authors define the signal-to-noise ratio (see Eq. (2.5) in that reference)} 
%\citep[see Example 1 in][]{step:zhu:22}, the  Kesten-Stigum (KS) threshold for possible recovery in the sparse regime is 
%\[
%SNR := \frac{\Big[\sum_{m=2}^M (m-1)\binom{n}{m-1} (\alpha^{(m)} - \beta^{(m)})/Q^{m-1} \Big]^2} {\sum_{m=2}^M (m-1)\binom{n}{m-1}\Big[ \frac{(\alpha^{(m)} - \beta^{(m)})}{Q^{m-1}} + \beta^{(m)}\Big] } 
%\]
%\blue{and conjecture that SNR$>1$ corresponds to the clusters recovery threshold.}
%\red{HERE, neglecting the fact that our $\pi$ is non-uniform, for A2 we have SNR =4 ; for A3 we have SNR = 4.7; for A3' we have SNR= 7.9; for B2 we have SNR=0.7 and SNR.abs= 3.7 ; for B3 we have SNR=0.12 and SNR.abs=1.3, where in SNR.abs we take similar formula but with absolute values of $|\alpha^{(m)}-\beta^{(m)}|$ }

%%%%%%
\paragraph*{Parameter estimation accuracy.}
We also evaluate the accuracy of parameter estimation. As the parameter values may be extremely small (see Section~\ref{sec_SM:synth_pram} in SM), we choose to compute  the Mean Squared Relative Error (MSRE) between the true parameters (in the full model) and the estimated values, both for the prior probabilities $\pi_q$ and the probabilities of hyperedge occurrence $B^{(m)}_{q_1, \ldots, q_m}$. Specifically, we compute the aggregated MSRE over all the components of $\theta$ using the following formula:
\[
MSRE = \frac 1 {n_{rep}} \sum_{i=1}^{n_{rep}} \Big\{ \sum_{q=1}^{Q-1}\Big(\frac{\hat \pi_q^{i}-\pi_q}{\pi_q}\Big) ^2 + \sum_{m=2}^M\sum_{q_1\le \dots \le q_m}\Big( \frac{\hat B_{q_1,\dots,q_m}^{i}-\Bqm}{\Bqm}\Big)^2 \Big\},
\]
where $(\hat \pi_1^i,\dots,\hat \pi_{Q-1}^i, \{\hat B_{q_1,\dots,q_m}^{i}\}_{m,q_1,\dots,q_m})$ is the set of free parameters estimated on the $i$-th dataset by the full model and $n_{rep}=50$ is the number of replicates.

%%% MSRE
\begin{figure}[htbp]
    \centering
    \includegraphics[]{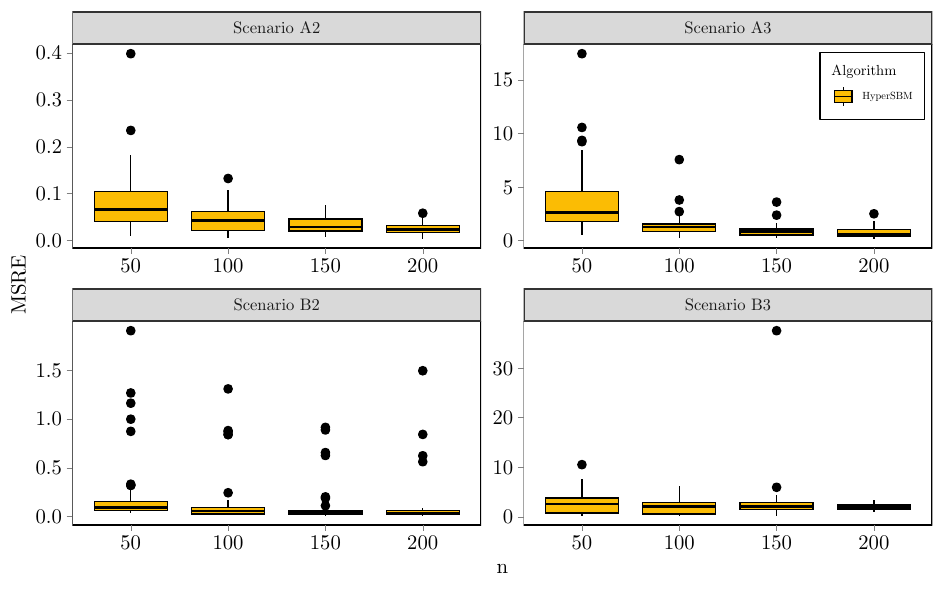}
    \caption{Boxplots of Mean Squared Relative Errors between true and estimated model parameters 
    for different settings X$Q$ (where X=A, B is the scenario and $Q=  2,3$ is the number of groups) and  number of nodes $n$ (along $x$-axis). First row (resp. first column) shows scenario A with communities (resp. $Q= 2$) while second row (resp. second column) shows scenario B with disassortative behaviour (resp. $Q= 3$).}
    \label{fig:mse_res}
\end{figure}

The corresponding results are summarized through the boxplots in Figure~\ref{fig:mse_res}. 
The relative errors are rather small, decreasing and showing a lower variance as the number of nodes increases. Note that the absolute values of MSRE cannot be compared between the cases $Q=2$ (first column) and $Q=3$ (second column), \blue{with very different scales on the $y$-axis}. Indeed, in the first case, the relative error is cumulated over a total of 1+3 +4=8 free parameters (in the full model), while this increases to 2+ 6+10=18 free  parameters when $Q=3$.

\subsection{Performance of model selection}
\label{sec:modsel}
In this section we assess the performance of ICL as a model selection criterion. 
%To this aim we simulate 50 hypergraphs under a Scenario A (communities) with $Q=3$ groups. The setting is denoted A3' to distinguish it from the former setting A3 above.
The simulated data is  fitted with our \texttt{HyperSBM} with a number of latent states ranging from 1 to 5. 

\begin{table}[htbp]
	\caption{Frequency (as a percentage) of the selected number of groups $Q$ for setting A3'.  
	Model selection is carried out by means of the ICL criterion. Results are computed over 50 samples for each value of $n$.} 
	\centering
	\begin{tabular}{C{1cm}C{3cm}C{3cm}C{3cm}C{3cm}}
		\toprule
		$ Q $   & $n=50$    	& $n=100$		& $n=150$	    & $n=200$	        \\
		\cmidrule{1-1}\cmidrule{2-3}\cmidrule{4-5}
		1		& $ 0\% $		& $ 0\% $       & $ 0\% $		& $ 0\% $           \\
		2		& $ 26\% $	    & $ 2\% $       & $ 0\% $		& $ 0\% $           \\
		3		& $ 74\% $	    & $ 98\% $	    & $ 100\% $	    & $ 100\% $         \\
		4		& $ 0\% $		& $ 0\% $       & $ 0\% $		& $ 0\% $           \\
		5		& $ 0\% $		& $ 0\% $       & $ 0\% $		& $ 0\% $           \\
		\bottomrule
	\end{tabular}
	\label{tab:mod_sel}
\end{table}

In Table~\ref{tab:mod_sel} we show the frequency of the selected number of groups for setting A3'. The correct model is selected in $ 74\% $ of cases for $ n = 50 $, in $ 98\% $ of cases for $ n = 100 $ and in $ 100\% $ of cases for $ n = 150, 200 $. We also compute the value of ARI of the classification obtained with 3 clusters depending on the selected number of latent groups. This value is always equal to 1 when the correct model is recovered, thus confirming the optimal behavior of \texttt{HyperSBM} already shown in Section~\ref{sec:sim_estim}.

\subsection{Line clustering through hypergraphs}
\label{sec:line}
Following \cite{Leordeanu12,kami:etal:19} and earlier references, we here explore the use of hypergraphs to detect line clusters of points in $\mathbb{R}^2$. Similarly to the construction of pairwise similarity measures, we  here resort on third order affinity measures to detect alignment of points since pairwise measures would be useless to detect alignment.  
Thus, for any triplet of points $\{i, j, k\}$, we use the mean distance to the best fitting line as a dissimilarity measure $d(i,j,k)$ and transform this through a Gaussian kernel to a similarity measure.  

\begin{figure}[ht]
    \centering
    \includegraphics[width=\textwidth]{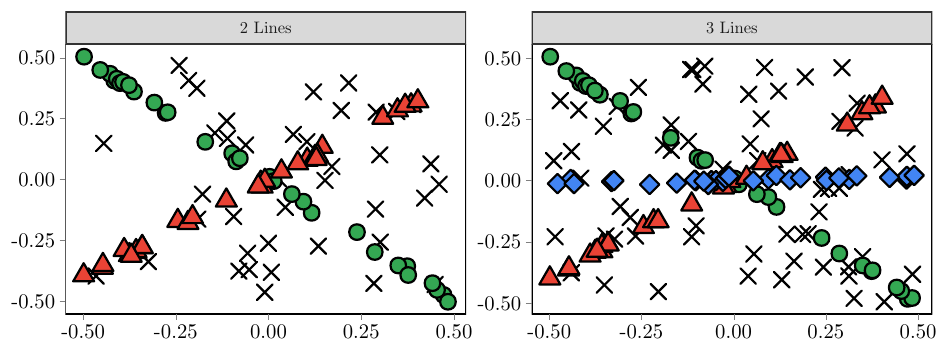}
    \caption{Sets of points from the line clustering experiments. Left: 2 lines (green dots and red triangles) plus noise (black crosses). Right: 3 lines (green dots,  red triangles and blue diamonds) plus noise (black crosses).}
    \label{fig:lines}
\end{figure}

We performed two different experiments, with either 2 or 3 lines. 
In each setting, we randomly generate the same number of points per line in the range $[-0.5, 0.5]^2$ and perturbed with centered Gaussian noise with standard deviation $0.01$. We then add noisy points, generated from uniform distribution on 
$[-0.5, 0.5]^2$.  
The particular settings of each  experiment are described in Table~\ref{tab:lines} and Figure~\ref{fig:lines} shows the resulting sets of points. 

\begin{table}[ht]
\caption{Description settings for the line clustering experiments.}
    \centering
    \begin{tabular}{c|C{3cm}C{3cm}C{3cm}C{3cm}}
    & \makecell{Number of points \\ per line} & \makecell{Number of \\ noisy points} & \makecell{Total number \\ of points} & \makecell{Mean number \\ of hyperedges}\\
    \hline
2 lines &30 &40 & 100& 1070.84\\    
3 lines & 30& 60 & 150&  587.70  
\end{tabular}
\label{tab:lines}
\end{table}

For both settings, we generated 100 different 3-uniform hypergraphs using the following procedure. We randomly selected 3 points $\{i, j, k\}$ and calculated the mean distance $d(i,j,k)$ to the best-fitting line. We then measured their similarity using a Gaussian kernel $\exp(-d(i,j,k)^2/\sigma^2)$ with $\sigma^2=0.04$. If the similarity was greater than a threshold $\epsilon=0.999$, we constructed a hyperedge $\{i,j,k\}$. This procedure resulted in both signal hyperedges, where all points belonged to the same line cluster, and noise hyperedges, where the points were sufficiently aligned without belonging to the same line. The signal-to-noise ratio of hyperedges was set to 2 for each hypergraph. We specifically simulated sparse hypergraphs, and the average number of hyperedges is presented in Table~\ref{tab:lines}. Additionally, isolated nodes in the hypergraph were excluded from the clustering analysis.

We applied our \texttt{HyperSBM} algorithm to cluster the nodes of these 3-uniform and sparse hypergraphs, and we compared the results with three different modularity-based approaches. The first two approaches, referred to as Chodrow\_Symm and Chodrow\_AON, are from \cite{chod:veld:bens:21} and are based on their general symmetric and all-or-nothing modularity, respectively. The third approach, referred to as Kaminski, is from \cite{kami:etal:19}. The modularity-based methods automatically select the number of groups, and for \texttt{HyperSBM}, we performed model selection using $Q\in\{1,\dots, 6\}$.

\begin{figure}[ht]
    \centering
    \includegraphics[width=\textwidth]{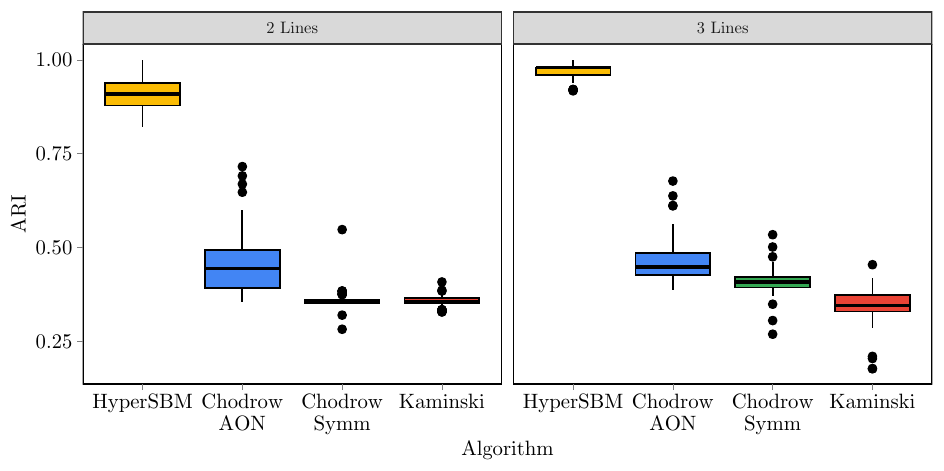}
    \caption{Boxplots of the ARI obtained by the different clustering methods on the line clustering problem. Left: 2 lines, right: 3 lines.}
    \label{fig:ARI_lines}
\end{figure}

\begin{figure}[ht]
    \centering
    \includegraphics[width=\textwidth]{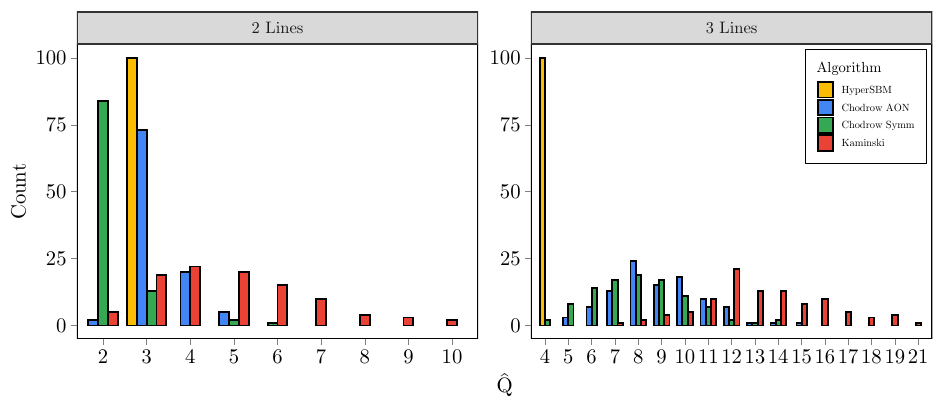}
    \caption{Estimated number of groups $\hat{Q}$ on the line clustering problem. Left: 2 lines  (true value of $Q$ is 3), right: 3 lines (true value of $Q$ is 4).}
    \label{fig:modelsel}
\end{figure}

Figure~\ref{fig:ARI_lines} displays the ARI obtained from the clustering results. We can observe that the modularity-based methods  fail to accurately recover the true original line clusters, resulting in lower ARIs. In contrast, \texttt{HyperSBM} shows good performance in this task, achieving higher ARIs. This difference in performance can be attributed to the tendency of modularity-based methods, especially the one by \cite{kami:etal:19}, to select a larger number of groups in this particular context, as evidenced in Figure~\ref{fig:modelsel}.

This experiment highlights the distinct behavior of \texttt{HyperSBM} compared to the modularity-based clustering methods, including the approach proposed by \cite{chod:veld:bens:21}, despite both methods being based on a Stochastic Block Model (SBM) framework with a maximum-likelihood approach. 

\section{Analysis of a co-authorship dataset}\label{sec:data}
\subsection{Dataset description}
We analyze a co-authorship dataset available at \url{http://vlado.fmf.uni-lj.si/pub/networks/data/2mode/Sandi/Sandi.htm}. The dataset originates from the bibliography of the book ``Product Graphs: Structure and recognition'' by Imrich and Klavz\u{a}r and is provided as a bipartite author/article graph.
To construct the hypergraph, following the approach of \cite{estr:rodr:06}, we consider authors as nodes and create hyperedges that link authors who have collaborated on the same paper. Further details regarding the dataset pre-treatment can be found in Section~\ref{sec_SM:dataset} of the SM, along with additional analyses.
In our analysis, we set $M=4$ and focused on the main connected component of the hypergraph, which consists of 79 authors and 76 hyperedges. Among these hyperedges, 68.5\% have a size of 2, while 29\% have a size of 3, and 2.5\% have a size of 4.

\subsection{Analysis with \texttt{HyperSBM}}
We conducted an analysis of this dataset using our \texttt{HyperSBM} package. The model selection based on the ICL criterion determined that there are 2 groups ($\hat Q=2$). One group consists of only 8 authors, while the remaining 71 authors belong to the second group.
Table~\ref{tab:nb_coauthors} displays the distribution of the number of distinct co-authors per author. Within the first group of 8 authors, 6 of them have the highest number of distinct co-authors, while the remaining 2 authors each have 4 distinct co-authors.
% Il numero di distinct co-authors dei nostri 8 sono: 10 12 10 11  8  4  7  4
%%%%

\begin{table}[ht]
    \caption{Distribution of the number of distinct co-authors per author. The first group contains the 6 authors having the largest number of distinct co-authors (between 7 and 12) plus 2 authors with 4 co-authors each.}
    \centering
    \begin{tabular}{l|ccccccccccc}
        \text{Nb co-authors}    & 1 & 2 & 3 & 4 & 5 & 6 & 7 & 8 & 10& 11& 12\\
        \hline
        \text{Count}        & 23 &  27&   13 &   6 &   2 &   2&    1&    1&     2&     1&     1 
    \end{tabular}
    \label{tab:nb_coauthors}
\end{table}

Coming back to the bipartite graph of authors and (co-authored) papers, we looked at the degree distribution of the authors, given in Table~\ref{tab:degree_authors}. This corresponds to the distribution of the number of co-authored papers per author. We observed that 5 of the 8 authors from our first group are the ones that co-published the most, the three others having also high degree (one of degree 5 and two of degree 4). %  nb of publications dei nostri 8 è: 8  7 10 13  5  4  6  4
Thus, our first group is made of authors (among) the most collaborative ones, which are also  (among) the most prolific ones. 

\begin{table}[ht]
    \caption{Degree distribution of authors in the bipartite graph. Our first group contains the 5 most collaborating authors, one of the sixth, plus 2 authors with degree equal to 4.}
    \centering
    \begin{tabular}{l|cccccccccccc}
        \text{Author degree}& 1 & 2 & 3 & 4 & 5 & 6 & 7 & 8 & 10& 13\\ 
        \hline
        \text{Count}      &44 &  14 &   6    & 6&     4 &    1 &    1    & 1&     1&     1
    \end{tabular}
    \label{tab:degree_authors}
\end{table}

Neither the first nor the second group inferred by \texttt{HyperSBM}  are communities. Indeed we obtained the following estimated values from the size-$2$ hyperedges: 
% 0.04204852 0.0516748 0.008178429
$\hat B_{11} \simeq 4.2\%$ is of the same order as $\hat B_{12}\simeq 5.1\%$ while $\hat B_{22}\simeq 0.8\%$ is around five times smaller. 
%(We have dropped from our notation $\hat B$ the index $n$ and only use the size $m$ of the hyperedges to distinguish the parameter values).
This means that the first group contains authors that have written with  authors from the two groups while the second group is made of authors who have less co-authored papers with people of their own group. 
Looking now at size-3 hyperedges, we get that 
%  0.0002222272 0.001843578 0.0007088518 5.882527e-05
$\hat B_{111}\simeq 2\cdot 10^{-4}$ ; 
$\hat B_{112}\simeq 18\cdot 10^{-4}$ ;
$\hat B_{122}\simeq 7\cdot 10^{-4}$ and 
$\hat B_{222}\simeq 0.6\cdot 10^{-4}$. 
The most important estimated frequency is $\hat B_{112}$ that concerns 2 authors of the small first group co-authoring a paper with one author of the large second group. The second most important estimated frequency is $\hat B_{122}$ and is  obtained for one author from small first group co-authoring a paper with two  authors of the large second group. The remaining frequencies of size-$3$ hyperedges are negligible. This characterizes further the first groups as being composed by authors that do co-author with their own group as well as with authors from the second one.   

Finally, looking now at size-4 hyperedges, 
% 3.619797e-242 2.46146e-163 1.248545e-84 4.267645e-06 4.040716e-131
the only non negligible estimated frequency is obtained  for $\hat B_{1222}\simeq 4\cdot 10^{-6}$. We note here that the \blue{frequencies $\hat B$'s with $m=3$ or $4$} are intrinsically on different scales, as \blue{also happens with $m=2$ or $3$}.
%are the quantities $B^{(2)}$'s and $B^{(3)}$'s.
So again, authors from group 1 co-authored with the others authors. (Note that the first group is not large enough for a \blue{size-4 $\hat B$} frequency with  at least 2 authors in that group 1 to be non negligible).

\subsection{Comparison with 2 other methods}
We first compared our approach with the hypergraph spectral clustering (HSC) algorithm proposed in~\cite{ghos:dukk:17:aos}. 
Let us recall that spectral clustering does not come with a statistical criterion to select the number of groups. 
Looking at the partition obtained with $Q=2$ groups, spectral clustering output groups with sizes 24 and 55, respectively. %We recall that  spectral clustering tends to output comparable sizes groups. 
These groups are neither characterized by the number of co-authors nor their degrees in the bipartite graph (see details in SM). 
Indeed, in our case the best clusters are not communities and their sizes are very different, while we recall that spectral clustering tends to: i) extract communities ; ii) favor groups of similar size.\\

We then analysed the same dataset as a bipartite graph of authors/papers with the \texttt{R} package \texttt{SBM}  through the function \texttt{estimateBipartiteSBM}  \citep{SBM}. 
This method infers a latent  blockmodel (that in fact corresponds to a SBM for bipartite graphs) and automatically selects a number of groups on both parts (authors and papers). 
\blue{The method relies on the same core \texttt{VEM} algorithm as ours, adapted to the bipartite graphs context.  }
Hereafter, we refer to this method as the \texttt{Bipartite-SBM} implementation. Let us underline here that while the bipartite stochastic blockmodel can be written as a particular case of a HSBM, the converse is not true (see Section~\ref{sec_SM:HSBM_not_bipSBM} in the SM). \blue{In particular, our hypergraph SBM is not constrained by the need to cluster the set of hyperedges.}

The \texttt{Bipartite-SBM} also selected 2 groups of authors (and 1 group of papers). There was one small group with 4 authors, entirely contained in our first small group; it corresponds to authors that have the highest degree in the bipartite graph and %also the 4 authors having 
the highest number of co-authors. 
So, \texttt{Bipartite-SBM} output a very small group of the most prolific and the most collaborative authors in this dataset. Further details about the distinctions between these groups and the ones obtained by  \texttt{HyperSBM} are given in SM. 

As a conclusion, we see that while the outputs of \texttt{Bipartite-SBM} and \texttt{HyperSBM} may seem close on this specific dataset, they are nonetheless different. On the other hand, and still on this specific dataset, the spectral clustering approach outputs results that are completely different from those of \texttt{HyperSBM}.

%%%%%%%%%%%%%%%%%%%%%%%
%%%%%%%%%%%%%%%%%%%%%%%
\section{Discussion}	
\label{sec:discuss}	

%\paragraph*{Distinction from previous proposals.}
%We first highlight the distinctive methodological aspects of our approach compared to other existing methods.
%
%\cite{ghos:dukk:14,ghos:dukk:17:aos} obtained error bounds that converge to zero only for specific cases of the \eqref{eq:aff-m} model with equally-sized groups. Their results are limited to these specific scenarios, whereas our approach extends beyond such constraints.
%
%References such as \cite{ke:shi:xia:20,ahn:lee:suh:18,chie:lin:wang:19} primarily focus on community detection, which means they only identify clusters that correspond to communities. In contrast, our method is not limited to detecting community-like structures, but can find clusters that may not align with traditional notions of communities (e.g. disassortative clusters).
%
%In Sections~\ref{sec:simus} and \ref{sec:data}, we provide numerical illustrations that highlight the differences between our algorithm and those proposed by \cite{chod:veld:bens:21, kami:etal:19, ghos:dukk:17:aos}, as well as the bipartite SBM.\\
	
\blue{We have proposed a hypergraph stochastic block model for simple hypergraphs and general clusters types, \emph{i.e.} our work is not limited to community detection and/or equally-sized clusters. This is in sharp contrast with most existing approaches. For example, \cite{ghos:dukk:14,ghos:dukk:17:aos} obtained error bounds that converge to zero only for the \eqref{eq:aff-m} model with equally-sized groups and assuming moreover that $\alpha^{(m)}>\beta^{(m)}$. Moreover, references such as \cite{ke:shi:xia:20,ahn:lee:suh:18,chie:lin:wang:19} primarily focus on community detection, which means they only identify clusters that correspond to communities. 
%%%%% Consistency and accuracy 
Our inference procedure is based on a maximum-likelihood approach, which should in principle provide some statistical guarantees. 
While consistency and asymptotic normality of the variational and the maximum likelihood estimators in our HSBM is left for future work, we believe that such results could be obtained following approaches used in the context of  graphs SBMs \citep{celisse2012_EJS,bickel2013_AOS}.
It is worth noting that while \cite{chod:veld:bens:21} initially employ a maximum likelihood approach, they deviate from that setting for their inference procedure. In contrast, our method retains the maximum likelihood framework throughout the inference process.
The maximum likelihood approach also enables the use of a penalized criterion for model selection. 
The SBM for hypergraphs presented in \cite{bala:21} is highly general. However, their least-squares estimator for a hypergraphon model is computationally infeasible. Additionally, their Algorithm 1 is dedicated to community detection and does not provide general cluster recovery.
}
	
%%%%%%
%\paragraph*{Generalizations.} 
Our model can accommodate self-loops without significant changes by allowing for $m=1$. Furthermore, it can be easily extended to handle \mmh\ (with or without self-loops) by incorporating a zero-inflated or deflated Poisson distribution on the conditional distribution of the hyperedges. In a more general setting, the conditional Bernoulli distribution can be replaced with any parametric distribution to handle weighted hypergraphs, and it could also easily incorporate covariates. This flexibility allows for the adaptation of our model to various types of hypergraph data.

%%%%%%%%
% Efficiency and submodels
\blue{While an important challenge is to reduce the complexity of our approach, some gain could be provided by constraining the parameter set. For instance, \cite{conti:etal:22} consider a Poisson HSBM, where the connectivity parameter is non-zero only between nodes in the same cluster. While this  assumption is quite restrictive, it is mitigated by the introduction of overlapping clusters. In the same way,  \cite{ruggeri:etal:23} propose a similar model where the connectivity parameter is the sum of nodes-pairs contributions, resulting in a model that differs from what could be obtained through a clique reduction graph (namely, the graph obtained from hyperedges  transformed into cliques). In both cases, these constraints on the parameters considerably reduce the complexity of the inference procedure which is based on a variational-like  approach (but does not rely on an evidence lower bound).  We believe that similar techniques could be useful in our case and plan to explore that in future works.}

%%%%%%%%%%%%%%%%%%%%%%%
%%%%%%%%%%%%%%%%%%%%%%%

%%%%%%%	
\section{Codes availability}
\label{sec:available}
The algorithm implementation in \texttt{C++} is available as an \texttt{R} package called \texttt{HyperSBM} at \url{https://github.com/LB1304/HyperSBM}. The Supplementary Material,  the files to reproduce the synthetic experiments and the dataset analysis are available at \\
\url{https://github.com/LB1304/Hypergraph-Stochastic-Blockmodel}. 

\section*{Acknowledgements}
Funding was provided by the French National Research Agency (ANR) grant ANR-18-CE02-0010-01 EcoNet. L. Brusa thanks the financial support from the grant ``Hidden Markov Models for Early Warning Systems'' of the University of Milano-Bicocca (2022-ATEQC-0031)

%%%%%%%%%%%%%%%%%%%%%%%%%%%%%%%
%%%%%%%%%%%%%%%%%%%%%%%%%%%%%%%
%%%%%%%%%%%%%%%%%%%	
%%%%%%%%%%%%%%%%%%%	
\appendix
\newpage

%%%%%%%%%%% SUPP MAT %%%%%%%%%%%%%%%%
%%%%%%%%%%%%%%%%%%%%%%%%%%%%%%%%%
%%%%%%%%%%%%%%%%%
%%%%%%%%%%%%%%%%%

\begin{center}
{\Large Supplementary Material to: Model-based clustering in simple hypergraphs through a stochastic blockmodel \\
By Luca Brusa \&  Catherine Matias
}
\end{center}

\vspace{1cm}

All non-alphabetic references are concerned with the main text.  

\section{Limits of the bipartite graphs representation of hypergraphs}
\label{sec:bipartite}
\subsection{Bipartite graphs and \mmh\ with self-loops equivalence}
Some early analyses of hypergraphs rely on the embedding of the former into the space of
bipartite graphs \citep[see for e.g.][]{batt:cenc:iaco:etal:20_SM}. 
%%%%
Indeed, any hypergraph $\mathcal{H}=(\mathcal{V}, \mathcal{E})$ where $ \mathcal{V}$ is the set of nodes
and $\mathcal{E}$ the set of hyperedges may be
represented as a bipartite graph with two parts. The top part is simply the set  $\mathcal{V}$ of hypergraph nodes, while the bottom part is the set $\mathcal{E}$ of  hyperedges 
and there is a link between $v \in \mathcal{V}$ and $e\in \mathcal{E}$ whenever
node $v$ belongs to hyperedge $e$ in the original hypergraph
$\mathcal{H}$.

\begin{figure}
\includegraphics[width=\textwidth]{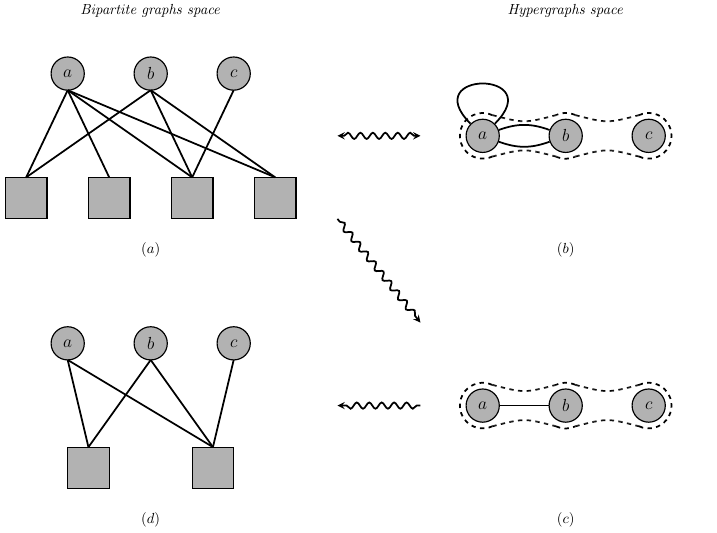}
\caption{(a) A bipartite graph $\mathcal{G}$; (b) Projection of $\mathcal{G}$ into the \mmh\ with self-loops space, choosing the top nodes as the new set of nodes. Hyperedges are $\{a\}, \{a,b\}, \{a,b\}, \{a,b,c\}$. The applications from (a) to (b) are invertible bijections, one being the inverse of the other; 
(c) Projection of $\mathcal{G}$ on the simple hypergraphs subspace. Hyperedges are $\{a,b\}, \{a,b,c\}$. (d) Embedding of the simple hypergraph in (c) in the bipartite graphs space. Note that (a) and (d) are not the same bipartite graph.}
\label{fig:not_bijection}
\end{figure}

Now, it is possible to define a ``converse'' application from bipartite graphs to hypergraphs. Indeed, any bipartite graph can be projected into two distinct hypergraphs, by choosing one of the two parts as the nodes set and forming a hyperedge with any set of nodes that are neighbors (in the bipartite graph) of the same node (belonging to the second part). A major difference  appears whether we consider simple hypergraphs or \mmh\ with self-loops. In \mmh\  (not to be confused with \mh) hyperedges may appear several time so that these are weighted hypergraphs with integer valued weights. We also allow for self-loops, \emph{i.e} hyperedges of cardinality 1. 
%%%%%
%Let us first recall that bipartite graphs are routinely projected into two distinct %(simple and undirected) graphs by choosing one of the two parts as the nodes set and forming an edge between any two nodes that share a neighbor in the bipartite graph. In the same way, 
Then, this application from bipartite graphs to hypergraphs slightly differs depending on whether we allow the image of a bipartite graph to be a \mmh\ with self-loops or a simple hypergraph. In the first case, all the information from the bipartite graph will be encoded in the \mmh\ with self-loops; while in the second case, part of the information will be lost. This is illustrated on a toy example in Figure~\ref{fig:not_bijection}.

%As a consequence, starting from a bipartite graph (Figure~\ref{fig:not_bijection}, (a)) and composing the two processes, namely projecting it on a simple hypergraphs space (Figure~\ref{fig:not_bijection}, (c)) and then embed the latter in the bipartite graphs space (Figure~\ref{fig:not_bijection}, (d)), the resulting bipartite graph is not the same as the original one. 
%Indeed, part of the information is lost in the process. 
The embedding of the simple hypergraphs space into the bipartite graphs space is not the inverse of the natural projection of bipartite graphs into simple hypergraphs. Thus, models of bipartite graphs are inappropriate to handle simple hypergraphs, as the former generally put mass on any bipartite graph, notwithstanding the fact that not all of these may be realized as the image of a simple hypergraph. 
%%
%As a consequence, this embedding spans a small part of the bipartite graphs space. 
%In other words, while any bipartite graph can produce two distinct hypergraphs with no specificities, a hypergraph produces a very particular type of  bipartite graph. For instance, a bipartite graph with nodes of degree one in its bottom part is not a hypergraph (at least not if we do not allow for self-loops in the hypergraph). Moreover, two nodes from the bottom part cannot share exactly the same neighbors in the top part (otherwise, it means these are the same hyperedges, which is not allowed unless we consider multiple hyperedges).  
For the same reason, preferential attachment models of bipartite graphs \citep{guil:lata:04} may not be directly used for simple hypergraphs as they would produce unconstrained bipartite graphs that do not necessarily come from simple hypergraphs. 
%Section~\ref{app:bipartite} from SM contains further considerations in the same line. 

\subsection{Artifacts induced by bipartite graphs models}
\label{app:bipartite}
%We give here additional considerations to the issues from Section~\ref{sec:bipartite}.

In order to view a bipartite graph as a hypergraph, one first needs to select the top and bottom parts. Swapping the role of the two parts will in general give another hypergraph. Statistical models of bipartite graphs handle the two parts symmetrically and do not differentiate between a top and a bottom part. They are thus inadequate for modeling hypergraphs. 

One may also note that most random bipartite graphs models are designed for fixed parts sizes, which induces, on top of a fixed number of nodes, a fixed number of hyperedges in the corresponding hypergraph model, an artifact which is not always desirable. For instance the uniformly random hypergraphs model allows for any possible density on the hyperedges. 

A last example of inadequacy is given by configuration models on bipartite graphs that induce configuration models on hypergraphs. In these models, the degree distributions in each part are kept fixed. When projected in  the hypergraphs  space, that means that the degrees of the nodes and the sizes of the hyperedges are kept fixed. Then, relying on shuffling algorithms to explore the space of this configuration model, one will loose the
labels on the bottom part (the hyperedges part) as these are
automatically induced by the new edges of the bipartite graph and the
labelling of the top part (the nodes part). As a consequence, if a specific node tends to take part in large size hyperedges, this information is lost in the configuration model issued from bipartite graphs. 

To our knowledge, there is no configuration model on hypergraphs that only keeps the nodes degrees sequence fixed. We mention that Section 4 from \cite{chod:20_SM} provides a discussion about the limitations  of the embedding approach in terms of the types of hypergraph null models from which we can conveniently sample. In particular, \cite{chod:20_SM} establishes that there is no obvious route for vertex-label sampling in hypergraphs through bipartite random graphs. 

%%%%%
\subsection{HSBM is not a bipartite SBM}
\label{sec_SM:HSBM_not_bipSBM}
In this section, we briefly outline that (i) while the  bipartite stochastic blockmodel can be seen as a  particular case of HSBM, (ii) the converse is not true in general. \blue{The main point here is that  hypergraph SBMs are more general than bipartite SBMs because they are not constrained by the assumption of the existence of a finite clustering on the hyperedges. }

To see point (i), let us consider a bipartite SBM on a graph $\mathcal{G}$ with nodes divided in 2 parts, say $\mathcal{V}=\{1,\dots,n\}$ and $\mathcal{U}=\{1,\dots,e\}$. The model has $Q$ groups (resp. $R$ groups) on the subset of nodes $\mathcal{V}$ (resp. $\mathcal{U}$), with group  proportions $\pi$ (resp. $\eta$). We let $Z_1,\dots, Z_n$ (resp. $W_1,\dots,W_e$) denote the latent groups of nodes $\mathcal{V}$ (resp. $\mathcal{U}$).

The model is also given by a connectivity matrix $M$ of size $Q\times R$ whose entries $M_{qr}$ are the conditional probabilities that a node in $\mathcal{V}$ from group $q$ connects a node in $\mathcal{U}$ from group $r$. In other words $M_{qr}=\mathbb{P}(X_{iu}=1|Z_i=q, W_u=r)$ where $X=(X_{iu})$ is the $n\times e$ incidence matrix of $\mathcal{G}$.

Now consider the hypergraph $\mathcal{H}$ constructed on the set of nodes $\mathcal{V}$ and whose hyperedges are obtained by looking at the set of nodes in $\mathcal{V}$ connected to a same node in $\mathcal{U}$. (A similar construction could be made with swapping the roles of $\mathcal{V}$ and $\mathcal{U}$). 
Then, the probability distribution of $\mathcal{H}$ under the induced bipartite SBM is exactly a HSBM with $Q$ groups, with group proportions $\pi$ and parameters 
\begin{align*}
\Bqm &= \mathbb{P}(\Ym=1|Z_{i_1}=q_1,\dots,Z_{i_m}=q_m) \\
&= \mathbb{P}(X_{i_1,u}=1, \dots , X_{i_m,u}=1 |Z_{i_1}=q_1,\dots,Z_{i_m}=q_m)\\
&= \sum_{r=1}^R \mathbb{P}(X_{i_1,u}=1, \dots , X_{i_m,u}=1, W_{u}=r|Z_{i_1}=q_1,\dots,Z_{i_m}=q_m)\\
&= \sum_{r=1}^R \eta_r \prod_{q=q_1}^{q_m} M_{qr}, 
\end{align*}
where $u$ is the node that connects $\{i_1,\dots,i_m\}$ into a hyperedge. 
So we see that the bipartite SBM induces a HSBM with constrained connection probabilities. \\

Let us now explain why (ii) the converse is not true in general. We start from a HSBM with $Q$ groups and parameters $(\pi, (\Bqm)_{q_1,\dots,q_m})_{2\le m \le M}$ on a hypergraph $\mathcal{H}$ with set of nodes $\mathcal{V}$. Considering $\mathcal{U}=\{1,\dots,e\}$ where $e$ is the number of hyperedges in $\mathcal{H}$, we construct a bipartite graph $\mathcal{G}$ with nodes $\mathcal{V}\times \mathcal{U}$ and links between any $i\in \mathcal{V}$ and any $u\in \mathcal{U}$ whenever node $i$ belongs to hyperedge $u$ in the hypergraph $\mathcal{H}$. Now, if there is a bipartite SBM on $\mathcal{G}$ with same distribution as our HSBM, then necessarily it has $Q$ groups on $\mathcal{V}$, with group proportions given by $\pi$. We let $R$ denote the number of groups on such a model on $\mathcal{U}$, together with $\eta$ the corresponding group proportions, and $M$ the $Q\times R$ matrix of connection probabilities. Then we observe that $\eta$ and $M$ should satisfy the relations:
\begin{equation}
    \label{eq:bipSBM_constraint}
\forall 2\le m\le M, \forall q_1,\dots, q_m \in \{1,\dots,Q\}^m, \quad 
\Bqm =\sum_{r=1}^R \eta_r \prod_{q=q_1}^{q_m} M_{qr}.
\end{equation}

Here, we first remark that the bipartite SBM fit on the co-authorship dataset (from Section~\ref{sec:data}) selected $R=1$, thus inducing hyperedges connectivity parameters with a product form 
\[
\Bqm = \prod_{q=q_1}^{q_m} M_{q1}.
\]
Our fitted HSBM on this same dataset did not result  in hyperedges connectivity parameters with a product form, which establishes that the models are clearly different. 

Now, more generally, we could ask whether for given parameters $(\Bqm)_{2\le m \le M}$, there exist some values of $R, \eta$ and $M$ such that~\eqref{eq:bipSBM_constraint} is satisfied. The answer is: not always. 
To see this, consider for instance $Q=2$ and remark the relation between the two quantities
\begin{equation}
    B^{(2,n)}_{11} =\sum_{r=1}^R \eta_r M_{1r}^2\qquad \text{ and } \qquad
    B^{(3,n)}_{111} =\sum_{r=1}^R \eta_r M_{1r}^3, \label{eq:mconstraint}
\end{equation}
so that $B^{(2,n)}_{11}$ and $B^{(3,n)}_{111}$ cannot be chosen independently. 

%%%%%%%%%%%%%%%%
\section{The need for simple hypergraphs models}
\label{sec:multi}
In this section, we discuss modeling differences between \mh\ where multiset hyperedges are allowed, versus  simple hypergraphs where hyperedges are subsets of nodes. 
We recall that \mh\ allow for repeated nodes in a same hyperedge, the latter being defined as a multiset of nodes.
Multiset hyperedges generalize in some sense the notion of self-loops in graphs and thus are a natural extension to consider. However, they are not appropriate in all contexts. For instance, a co-authorship dataset cannot contain hyperedges with repeated nodes (but may contain a self-loop of a unique author). In the same way, a social interaction hypergraph does not contain multisets hyperedges; triadic interactions are restricted to 3  different individuals and self-loops are not allowed. 
In the meantime, they are natural in other contexts; consider, \emph{e.g.}, chemical reaction hypergraphs where the multiplicity plays the role of the stoichiometric coefficients \citep{flam:barb:stad:15}. We first argue that multisets-hypergraph models are inappropriate for analysing simple hypergraphs.

\subsection{A motivating example}
\label{sec:motivation}
Let us first focus on parameter estimates rather than on clustering. 
For the sake of simplicity, we restrict our attention to $3$-uniform hypergraphs on a set of $n$ nodes and consider two different models. The first one, denoted as MH, acts on $3$-uniform \mh\ and draws a hyperedge between any 3 nodes, not necessarily distinct, with  probability $p_{\text{MH}}$. The second one, denoted as SH, acts on $3$-uniform simple hypergraphs and draws a hyperedge between any 3 distinct nodes with  probability $p_{\text{SH}}$.

Now, we consider a toy example of observing a simple hypergraph $\mathcal{H}$ with $n=3$ nodes and only one hyperedge $e=\{1,2,3\}$. This dataset could correspond to observing for instance one publication with 3 authors. 
When analysed under the MH model, the density of our observed hypergraph is estimated by 
\[
\hat p_{\text{MH}} = 1/ 27
\]
because there are $n^3= 27$ possible size-$3$ multiset hyperedges under this model, and just one of these is observed. On the contrary, when analysed under the SH model, we infer a density of 
\[
\hat p_{\text{SH}} = 1
\]
because the only possible size-$3$ hyperedge is observed. As a consequence, the statistical conclusions drawn on this dataset will highly differ depending on whether we restrict attention to simple hypergraphs or work with more general \mh. This choice of the ambient space has to be made according to the specificities of the dataset. 
This simple and elementary example shows that it is not possible to statistically analyse a simple hypergraph with a \mh\ model without erroneous conclusions.

%%%%%%%
%\subsection{Computational challenge in the simple hypergraph case}
\subsection{Null model for \mh\ in modularity computations}
\label{sec:mh}
The main technical difference between \mh\ and simple hypergraphs analysis comes from the enumeration of size-$m$ subsets of nodes. 
In the \mh\ setting, the summations over multisets of nodes $\{i_1, \dots, i_m\} \in \{1,\dots, n\}^m$ factorize into $m$ independent sums. On the contrary, in the simple hypergraph setting, the summations involve sets of nodes  $\{i_1, \dots, i_m\}$ that are constrained to be distinct. As a consequence, 
such a factorization is impossible. 

Let us consider a  concrete example. We already emphasized the fact that modularity criteria for hypergraphs have been proposed only in the \mh\ setting \citep{kami:etal:19_SM,chod:veld:bens:21_SM}. 
Modularities are generally constructed as deviation measures of the number of hyperedges from their expected number under a null model. For instance in the graphs context, the Newman and Girvan modularity of a partition $(C_1,\dots, C_Q)$ of the nodes into $Q$ clusters is computed  as 
\begin{align*}
    \text{Modularity}(C_1,\dots, C_Q) &= \frac{1}{2|E|} \sum_{q=1}^Q \sum_{i,j \in C_q} \Big(A_{ij} -\frac{d_id_j}{2|E|} \Big) \\
    &= \frac 1 {2|E|} \sum_{q=1}^Q \sum_{i,j \in C_q} A_{ij} - \frac 1 {2|E|} \sum_{q=1}^Q \sum_{i,j \in C_q} \frac{d_id_j}{2|E|}, 
\end{align*}
where 
$A=(A_{ij})_{i,j}$ is the graph adjacency matrix, $d_i$ is the degree of node $i$, and $2|E|=\sum_{i} d_i$ is twice the number of edges. 
While the first part of these criteria enumerates only the occurring hyperedges, a quantity that is small in general as most hypergraph datasets are sparse, the second part needs to account for all subset of nodes in the graph (or at least in the same group $C_q$). In the case of graphs allowing self-loops,  this second term factorizes to 
\[
\frac 1 {2|E|} \sum_{q=1}^Q \sum_{i,j\in C_q} \frac{d_id_j}{2|E|} = \frac 1 {2|E|}\sum_{q=1}^Q \frac{(\sum_{i\in C_q} d_i)(\sum_{j\in C_q} d_j)}{2|E|}= \sum_{q=1}^Q\frac{\text{Vol}(q)^2}{(2|E|)^2} ,  
\]
where the computation of the volume $\text{Vol}(q)=\sum_{i\in C_q}d_i$ has time complexity of $O(n)$. Similarly, for \mh\ the modularity computed in  \cite{chod:veld:bens:21_SM} uses two main terms: the first is a cut term that depends only on occurring hyperegdes while the second relies on volumes of latent configurations of the nodes \citep[see Eq.~(12) and (13) in][]{chod:veld:bens:21_SM}. 
On the contrary, in the simple hypergraph setting, enumerating all subsets of nodes constrained to be distinct requires enumerating  \[
\sum_{m=2}^M \binom{n}{m}
\]
elements for a hypergraph with $n$ nodes and maximum hyperedge size $M$. This quantity is huge and represents the main computational limit when analysing hypergraphs (our approach to this issue is detailed in Section~\ref{app:details} from SM).

%%%%%%%%%%%%
%%%%%%%%%%%%
\section{Proof of Theorem 2}
\label{app:ident} 

The proof closely follows the structure of the proof for Theorem 2 in \cite{allm:etal:11_SM} for the SBM on simple graphs, with the generalization to simple $m$-uniform HSBM. 
For the sake of completeness, we provide here the complete proof of Theorem~\ref{thm:ident}.
Note that the key element that distinguishes our proof from the one in \cite{allm:etal:11_SM} is clearly identified below.

\paragraph{The strategy relying on Kruskal's result.}
The proof strongly relies on an algebraic result from \cite{krus:77_SM} that appeared to be a powerful tool to establish
identifiability results in various models whose common feature is the presence of discrete latent groups and at least
three conditionally independent random variables.
We first rephrase Kruskal's result in a statistical context.  Consider
a latent random variable $V$ with state space $\{1,\ldots,r\}$ and
distribution given by the column vector $\mathbf v =(v_1,\ldots,
v_r)$. Assume that there are three observable random variables $U_j$
for $ j=1,2,3$, each with finite state space
$\{1,\ldots,\kappa_j\}$. The $U_j$s are moreover assumed to be
independent conditional on $V$.  Let $M_j$, $j=1,2,3$ be the
stochastic matrix of size $r\times \kappa_j$ whose $i$th row is
$\mathbf{m}_{i}^j = \PP(U_j=\cdot \mid V=i)$.  Then consider the $3$-dimensional array (or tensor) with dimensions 
 $\kappa_1\times\kappa_2\times \kappa_3$ denoted  $[\mathbf
v;M_1,M_2,M_3]$ and whose $(s,t,u)$ entry (for any $ 1\le s\le \kappa_1 , 1\le t\le \kappa_2 ,
1\le u\le \kappa_3 $) is defined by
\begin{align*}
  [\mathbf v;M_1,M_2,M_3]_{s,t,u} &=\sum_{i=1}^r v_i m^1_i(s) \,
                                    m^2_i(t) \, m^3_i(u) \\
  &=\sum_{i=1}^r \PP(V=i)\PP(U_1=s|V =i)\PP(U_2=t|V =i)\PP(U_3=u|V =i)\\
  & =\mathbb{P}(U_1=s,U_2=t,U_3=u) . 
\end{align*}
Note that $[\mathbf v;M_1,M_2,M_3]$ is left unchanged by
simultaneously permuting the rows of all the $M_j$ and the entries of
$\mathbf v$, as this corresponds to permuting the labels of the latent
classes.
Knowledge of the distribution of $(U_1,U_2,U_3)$ is equivalent to
knowledge of the tensor $[\mathbf v; M_1,M_2,M_3]$.\\
%%%%
Now, the \emph{Kruskal rank} of a matrix $M$, denoted $\rank_K M$, is the largest
number $I$ such that \emph{every} set of $I$ rows of $M$ are
independent.
%Note that this concept would change if we replaced``row'' by ``column,'' but we only use the row version in this
%article.
Note that for any matrix $M$, its  Kruskal rank is necessarily less than its rank, namely $\rank_K M\le \rank M$, 
and equality of rank and Kruskal rank does not hold in general.
However, in the particular case when a matrix $M$ of size $p\times q$
has rank $p$, it also has Kruskal rank $p$.
Now,  let $I_j=\rank_K M_j$. \cite{krus:77_SM} established the following result. If 
\begin{equation}
  \label{eq:kruskal_condition}
I_1+I_2+I_3\ge 2r+2,  
\end{equation}
then the tensor $[\mathbf v;M_1,M_2,M_3]$ uniquely determines $\mathbf v$ and the $M_j$, up to
simultaneous permutation of the rows. In other words, the set of
parameters $\{(\mathbf v, \mathbb{P}(U_j=\cdot \mid V))\}$ is uniquely
identified, up to label switching on the latent groups, from the distribution of the random
variables $(U_1,U_2,U_3)$.

Now, to obtain generic identifiability, it is sufficient to exhibit a single parameter value for which
\eqref{eq:kruskal_condition}  is satisfied. Indeed, the set of parameter values for which $\rank_K M_j$ is fixed can be
expressed through a Boolean combination of polynomial inequalities ($\neq$, or rather non-equalities) involving matrix minors in
those parameters. In the same way, the converse condition of \eqref{eq:kruskal_condition}, namely inequality
$I_1+I_2+I_3\le 2r+1$ is the finite Boolean combination of  polynomial non-equalities on the model parameters. This
means that this set of parameters is an algebraic variety. But an algebraic variety can only be either the whole
parameter space (in which case exhibiting a single value where \eqref{eq:kruskal_condition}  is satisfied would not be
possible) or a proper subvariety, thus a subspace of dimension strictly lower than that of the whole parameter space.

The strategy of the proof for showing identifiability of certain discrete latent class models developed in
\cite{allm:etal:11_SM} and other papers by the same authors is to embed these models in the context of Kruskal's
result just described. Applying Kruskal's result to the embedded
model, the authors derive partial identifiability results on the embedded model, and
then, using details of the embedding, relate these to the original model.

\paragraph*{Embedding the HSBM into Kruskal's setup.}
For some number of nodes $n$ (to be specified later), we let  $V=(Z_1,Z_2,\dots,Z_{n})$ be the latent random
variable, with state space $\{1,\ldots, Q\}^{n}$ and denote by $\mathbf{v}$ the corresponding vector of
its probability distribution.  The entries of $\mathbf{v}$ are of the form $\pi_1^{n_1}\cdots\pi_Q^{n_Q}$ for some
integers  $n_q \ge 0$ and such that $\sum_q n_q =n$.
We fix $m\ge 2$ and consider simple $m$-uniform hypergraphs on the set of nodes $\V=\{1,\dots, n\}$. Recall that
$\V^{(m)}$ is the set of all distinct $m$-tuples of nodes in $\V$ and $\{\Ym ; \{i_1,\dots, i_m\}\in
\V^{(m)}\}$ the set of all indicator variables corresponding to possible (simple)
hyperedges of a $m$-uniform hypergraph over $\V$. 
Now, we will construct below subsets $H_1,H_2,H_3 \subset \V^{(m)}$ of distinct $m$-tuples of nodes such that $H_i\cap H_j
=\emptyset$ for any $i\neq j$. Then, we choose the 3 observed variables $U_j$ ($1\le j\le 3$)  as the vectors of indicator
variables $U_j=(\Ym)_{\{i_1,\dots, i_m\}\in H_j}$. This induces  that $\kappa_j=2^{|H_j|}$ (where $|H_j|$ is the
cardinality of $H_j$). As the subsets $H_1,H_2,H_3$ do not share any $m$-tuple of nodes,
the random variables $U_j$ are conditionally independent given $V$. We are in the statistical context of Kruskal's result.

The goal is now to construct the 3 subsets $H_j$ of $m$-tuples such that their pairwise intersections are empty and such that
condition \eqref{eq:kruskal_condition} is satisfied (for at least one parameter value of the embedded model and thus
generically for this embedded model). This construction of the $H_j$'s proceeds in two steps: the base case and an
extension step.

Starting with a small set $\V_0=\{1,\dots, n_0\}$ of nodes, we define a matrix $A$ of dimension $Q^{n_0} \times
2^{\binom {n_0} m}$. Its  rows are indexed by latent configurations $v \in
\{1,\ldots, Q\}^{n_0}$ of  the nodes in $\V_0$, its 
columns  by the set of all possible states of the vector of indicator variables  $(\Ym)_{\{i_1,\dots, i_m\}\in \V_0}$, and the 
entries of $A$ give the probability of observing the specified states of the vector of indicator variables,
conditioned on the latent configurations $v$. Thus each column index corresponds to a
different simple $m$-uniform hypergraph on $\V_0$.
% We therefore refer to a column index as a \emph{configuration}.
The base case consists in exhibiting a value of ${n_0}$ such that this matrix $A$ generically has full row rank.
Then, in an extension step, relying on $n=n_0^2$ nodes, we construct the subsets $H_1,H_2,H_3$ with the desired
properties (namely their pairwise intersections are empty and \eqref{eq:kruskal_condition} is generically satisfied).

From Kruskal's theorem, we obtain that the vector $\mathbf{v}$ and the matrices $M_1,M_2,M_3$ are generically uniquely determined, up to simultaneous permutation of the
rows from the distribution of a simple $m$-uniform HSBM.

With these embedded parameters $\mathbf{v},M_1,M_2,M_3$ in hand, it is still necessary to recover the initial parameters
of the simple $m$-uniform HSBM: 
the group proportions $\pi_q$ and the connectivity matrix $B^{(m)}=(\Bqm)_{1\le q_1\le \dots \le q_m \le Q}$. This will be done in the conclusion.

\paragraph*{Base case.}
In the following, 
%we drop the exponents $(m,n)$ in the notation for the connection probabilities $B$ and simply 
we let $\bar B_{q_1,\dots , q_m}
=1 -{B}_{q_1,\dots , q_m}$.
%%%
The initial step consists in finding a value of ${n_0}$
such that the matrix $A$ of size $Q^{n_0}\times 2^{\binom {n_0} m}$
containing the probabilities of any simple  $m$-uniform hypergraph over these ${n_0}$ nodes, conditional on the hidden
node states,  generically has full row rank. 

The condition of having full row rank can be expressed as the
non-vanishing of at least one $Q^{n_0} \times Q^{n_0}$ minor of $A$. Composing
the map sending the parameters $\{B_{q_1,\dots, q_m}\} \to A$ with this collection of minors gives polynomials
in the parameters of the model.  To see that these polynomials are
not identically zero, and thus are non-zero for generic parameters,
 it is enough to exhibit a single choice
of the $\{B_{q_1,\dots, q_m}\} $ for which the corresponding matrix $A$ has full row rank.
%%%
We choose to consider parameters $\{B_{q_1,\dots, q_m}\}$ of the form
\[
  B_{q_1,\dots, q_m}=\frac{s_{q_1}s_{q_2}\dots s_{q_m}}{s_{q_1}s_{q_2}\dots s_{q_m}+t_{q_1}t_{q_2}\dots t_{q_m}}, \text{
  so } 
\bar {B}_{q_1,\dots , q_m}=\frac{t_{q_1}t_{q_2}\dots t_{q_m}}{s_{q_1}s_{q_2}\dots s_{q_m}+t_{q_1}t_{q_2}\dots t_{q_m}},
\]
with $s_q,t_l>0$ to be chosen
later.  However, since the property of having full row rank is unchanged
under non-zero rescaling of the rows of the matrix $A$, and all
entries of $A$ are monomials with total degree $\binom {n_0} m$ in
$B_{q_1,\dots, q_m},\bar{B}_{q_1,\dots, q_m}\}$, we may simplify the entries of $A$ by 
removing denominators, and consider the matrix (also called
$A$) with entries in terms of $B_{q_1,\dots, q_m}=s_{q_1}s_{q_2}\dots s_{q_m}$ and $\bar{B}_{q_1,\dots, q_m}=t_{q_1}t_{q_2}\dots t_{q_m}$. 

The rows of $A$ are indexed by the composite node states
$v \in \{1,\ldots,Q\}^{n_0}$, while its columns are indexed by
the $m$-uniform hypergraphs  $\mathcal{H}=(\ym)_{\{i_1,\dots, i_m\}\in \V_0}  \in \{0,1\}^{\binom {n_0} m}$.  For any composite hidden state $v 
\in \{1,\ldots,Q\}^{n_0}$ and any node $i \in \{1,\ldots,{n_0}\}$, let
$v(i) \in \{1,\ldots,Q\}$ denote the state of node $i$ in
the composite state $v$.  With our particular choice of the
parameters $B_{q_1,\dots, q_m}$, the $(v,\mathcal{H})$-entry of $A$
is given by
\begin{align*}
  \prod_{\{i_1\dots i_m\}\in \V_0^{(m)}} B_{v(i_1),\dots,v(i_m)}^{\ym}\bar{B}_{v(i_1),\dots,v(i_m)}^{1-\ym} =
   \prod_{i=1}^{n_0} s_{v(i)}^{d_i}
  t_{v(i)}^{{n_0}-1-d_i} ,
\end{align*}
where
\[d_i= \sum_{\substack{\{i_1\dots i_m\}\in \V_0^{(m)} \\ i \in \{i_1\dots i_m\}}}\ym
\]
is the degree of node $i$ in the
hypergraph $\mathcal{H}=(\ym)_{\{i_1,\dots, i_m\}\in \V_0} $.  With this choice of parameters $\{B_{q_1,\dots, q_m}\}$,  the
entries in a column of $A$ are entirely determined by the degree sequence $\mathbf{d}=(d_i)_{1\le
  i \le {n_0}}$ of the hypergraph under consideration. 
Two different hypergraphs may result in the same degree sequence, thus the same values in the two columns of $A$. 
For any degree sequence $\mathbf{d}=(d_i)_{1\le i \le {n_0}}$ arising
from a simple $m$-uniform hypergraph on $n_0$ nodes, let $A_{\mathbf{d}}$ denote a
corresponding column of $A$.
%%%
In order to prove that the matrix $A$ has full row rank, it is enough
to exhibit $Q^{n_0}$ independent columns of $A$. To this aim, we introduce polynomial functions whose independence is
equivalent to that of corresponding columns.

For each node  $i\in \{1,\ldots,{n_0}\}$ and each latent group $q\in
\{1,\ldots,Q\}$, introduce an indeterminate $X_{i,q}$ and a $Q^{n_0}$-size row
vector $\mathbf{X}=(\prod_{1\le i\le {n_0}}
X_{i,v(i)})_{v \in \{1,\ldots,Q\}^{n_0}}$. For each
degree sequence $\mathbf{d}$, we have 
\[
  \mathbf{X}A_{\mathbf{d}}= \sum_{v\in \{1,\ldots,Q\}^{n_0}}
  \prod_{1\le i\le {n_0}} s_{v(i)}^{d_i}
  t_{v(i)}^{{n_0}-1-d_i} X_{i,v(i)} 
= \prod_{1\le
    i\le {n_0}} \left(  s_{1}^{d_i}  t_{1}^{{n_0}-1-d_i}  X_{i,1} + \cdots +
s_{Q}^{d_i}  t_{Q}^{{n_0}-1-d_i}  X_{i,Q} \right).
\]
%To verify this,  notice that each monomial $(s^{d_1}_{q_1} t^{{n_0}-1-d_1}_{q_1} X_{1,q_1}) \cdots (s^{d_{n_0}}_{q_{n_0}} t^{{n_0}-1-d_{n_0}}_{q_{n_0}} X_{{n_0},q_{n_0}})$ obtained from multiplying out the product on the right corresponds to a choice of node states $q_i$ for nodes $i$, and hence a vector $v = (q_1, \dots, q_{n_0})$. Moreover, we obtain one such summand for each $v$.
%%%
Now, independence of a set of columns $\{A_\mathbf{d}\}$ is equivalent to
the independence of the corresponding set of polynomial functions $\{
\mathbf{X}A_{\mathbf{d}}\}$ in the indeterminates $\{X_{i,q}\}$. 
For a set $\mathcal{D}$ of degree sequences, to prove that the
polynomials $\{\mathbf{X}A_{\mathbf{d}}\}_{\mathbf{d}\in \mathcal{D}}$
are independent, we assume that there exist scalars $a_{\mathbf{d}}$
such that
\begin{equation}\label{eq:null_sum}
  \sum_{\mathbf{d}\in \mathcal{D}} a_{\mathbf{d}}
  \mathbf{X}A_{\mathbf{d}} \equiv 0,
\end{equation}
and show that necessarily all $a_{\mathbf{d}}=0$. 
This will be given by the following lemma from \cite{allm:etal:11_SM}. This lemma is originally formulated for a set 
$\mathcal{D}$ of degree sequences. However it is not specific to degree sequences;  it
applies for any sets $\mathcal{D}$ of sequences of integers indexed by $\{1,\dots,n_0\}$ and thus we phrase it in this
way. We  refer to \cite{allm:etal:11_SM} for its proof.

\begin{lemma}(Lemma 18 in \cite{allm:etal:11_SM}.) \label{lem:annihilate}
  Assume $n_0 \ge Q$.
  Let $\mathcal{D}$ be a set of $n_0$-length integer sequences such that for each
   $i\in \{1,\ldots,n_0\}$, the set of $i$-th coordinates $\{d_i 
  ~|~\mathbf{d}\in \mathcal{D}\}$ has cardinality at most $Q$. Then
  for generic values of $s_q,t_l$, for each $i$ and each $d^\star \in
  \{d_i~|~ \mathbf{d}\in \mathcal{D}\}$ there exist values of the
  indeterminates $\{X_{i,q}\}_{1\le q\le Q}$ that annihilate all the
  polynomials $\mathbf{X}A_{\mathbf{d}} $ for $\mathbf{d}\in
  \mathcal{D}$ except those for which $d_i=d^\star$.
\end{lemma}

The next step is to construct a set $\mathcal{D} $ of $n_0$-length integer sequences that satisfies
\begin{itemize}
\item for each
   $i\in \{1,\ldots,n_0\}$, the set of $i$-th coordinates $\{d_i 
  ~|~\mathbf{d}\in \mathcal{D}\}$ has cardinality at most $Q$ (condition in  Lemma~\ref{lem:annihilate}); 
  \item any $\mathbf{d} \in \mathcal{D} $ may be the degree sequence of a simple
$m$-uniform hypergraph;
  \item $|\mathcal{D}|\ge Q^{n_0}$.
  \end{itemize}  
With such a set at hand, by choosing one column of $A$ associated to each degree sequence in $\mathcal{D}$,
we obtain a collection of $|\mathcal{D}| \ge Q^{n_0}$ different columns of $A$. These
columns are independent since for each sequence $\mathbf{d}^\star\in
\mathcal{D}$, by Lemma~\ref{lem:annihilate} we can choose values of the
indeterminates $\{X_{i,q}\}_{1\le i\le n_0, 1\le q\le Q}$ such that all
polynomials $\mathbf{X}A_{\mathbf{d}}$ vanish, except
$\mathbf{X}A_{\mathbf{d}^\star}$, leading to $a_{\mathbf{d}^\star}=0$
in equation \eqref{eq:null_sum}. Thus, exhibiting such a set $\mathcal{D}$ is the last step to prove that $A$ has
generically full row rank.

\paragraph*{A construction specific to the hypergraphs case.}
Now, this is where our proof strongly differs from the one of Theorem 2 in
\cite{allm:etal:11_SM}. Indeed, the characterizations of degree sequences for graphs and simple $m$-uniform hypergraphs are completely
different. 
%Relying on a result by \cite{behr:etal:13}, we have exhibited such a set in Lemma~\ref{lem:setD}. 

%A crucial element in the proof involves constructing a set of degree sequences with specific properties. Due to the fundamental differences in characterizing which integer sequences can be realized as degree sequences between graphs and $m$-uniform hypergraphs, our construction differs from the one presented in the proof of Theorem 2 in \cite{allm:etal:11}.

Consider the following set of integer-valued sequences
\[
  \mathcal{D} =\Big\{ \mathbf{d}=(d_1,\ldots,d_{n_0}) ~|~ \text{ for } 1 \le i\le n_0, d_i \in
  \{m,2m, 3m, \ldots,Qm\} \Big\}.
\]

\begin{lemma}\label{lem:setD}
The set $\mathcal{D} $ of $n_0$-length integer sequences  satisfies 
\begin{itemize}
\item [(i)] for each
   $i\in \{1,\ldots,n_0\}$, the set of $i$-th coordinates $\{d_i 
  ~|~\mathbf{d}\in \mathcal{D}\}$ has cardinality at most $Q$; 
  \item [(ii)] For large enough $n_0$ (depending on $Q,m$), any $\mathbf{d} \in \mathcal{D} $ is  the degree sequence of a simple
$m$-uniform hypergraph over $n_0$ nodes;
  \item [(iii)] $|\mathcal{D}|\ge Q^{n_0}$.
\end{itemize}
\end{lemma}

Note that conditions $(i),(iii)$ imply that $\{d_i 
  ~|~\mathbf{d}\in \mathcal{D}\}$ should have cardinality exactly $Q$ and that $|\mathcal{D}|= Q^{n_0}$.

\begin{proof}[Proof of Lemma~\ref{lem:setD}]
Points $(i),(iii)$ are a consequence of the definition of $\mathcal{D}$. 
For any integer sequence $\mathbf{d}$, a necessary condition for $\mathbf{d}$ to be a degree sequence of a simple $m$-uniform
hypergraph over $n_0$ nodes is that  $m$ divides $\sum_{i} d_i$. Here, we
rather need sufficient conditions in order to prove $(ii)$. We rely on Corollary 2.2 in \cite{behr:etal:13_SM}.

\begin{corbehr}
  Let $\mathbf{d}$ be an integer-valued sequence with maximum term $\Delta$ and let $p$ be an integer such that
  $\Delta \le \binom{p-1}{m-1}$. If $m$ divides $\sum_i d_i$ and $\sum_i d_i\ge (\Delta-1)p+1$ then $\mathbf{d}$ is the degree sequence of a simple $m$-uniform
hypergraph. 
\end{corbehr}

Fix some $\mathbf{d} \in \mathcal{D}$. Note that by construction, $m$ divides $\sum_i d_i$. Let $\Delta$ be the maximum
value of this sequence and note that $\Delta \le Qm$. Thus we choose 
$p$ an integer such that $Qm \le \binom{p-1}{m-1}$.
Moreover, $\sum_i d_i \ge m n_0$ and $(\Delta - 1)p + 1 \leq \Delta p \le Qm p$. Then by choosing $n_0 \ge Qp$, we
obtain the desired result. 
\end{proof}

With Lemma~\ref{lem:setD} at hand, we have exhibited a set $\mathcal D$ with the desired properties, and thus established that $A$ has
generically full row rank.

\paragraph*{The extension step.}

The extension step builds on the base case, in order to
construct a larger set of $n=n_0^2$ nodes and subsets $H_1,H_2,H_3 \subset \V^{(m)}$ of distinct $m$-tuples of nodes in $\V=\{1,\dots,n\}$ with the desired properties. 
This step was first stated as Lemma 16 in \cite{allm:etal:09} in the context of simple graphs SBM and we extend it
below to our case. 

Let us recall that we want to construct  $H_1,H_2,H_3 \subset \V^{(m)}$ that are pairwise disjoint. 
 Then, with notation from above, we choose the 3 observed variables $U_j$ ($1\le j\le 3$)  as the vectors of indicator variables $U_j=(\Ym)_{\{i_1,\dots, i_m\}\in H_j}$.
 As the subsets $H_1,H_2,H_3$ do not share any $m$-tuple of nodes,
the random variables $U_j$ are conditionally independent given $V=(Z_1,\dots,Z_n)$. We let $M_j$ denote the $Q^n\times
2^{|H_j|}$ matrix of conditional probabilities of $U_j$ given $Z$.

\begin{lemma} \label{lem:blockinduct}
Suppose that for some number of nodes $n_0$, the matrix $A$ of size $Q^{n_0} \times 2^{\binom{n_0}{m}}$ defined above has
generically full row rank.  Then with $n=n_0^2$ there exist
  pairwise disjoint subsets $H_1,H_2,H_3\subset \V^{(m)}$ of $m$-tuples of nodes in $\V=\{1,\dots,n\}$ such that  for each $j$ the
  $Q^n\times 2^{|H_j|}$ matrix $M_j$  has generically full row rank ($Q^n$).
\end{lemma}

\begin{proof}[Proof of Lemma~\ref{lem:blockinduct}]
Let us describe the construction of $H_j$. We will partition the $n_0^2$
  nodes into $n_0$ groups of size $n_0$ in three different ways, each way leading to one $H_j$. Then each $H_j$ will be the
  union of the $n_0$ sets of all $m$-tuples made of some $n_0$ nodes.  
  Thus each $H_j$ has cardinality $n_0 \binom {n_0} m$.

Labeling the nodes by $(u,v)\in\{1,\cdots,n_0\}\times \{1,\cdots,n_0\}$, we picture the nodes as lattice points in a square grid. We take as the partition leading to
$H_1$ the rows of the grid, as the partition leading to $H_2$ the
columns of the grid, and as the partition leading to $H_3$ the
diagonals. In other words, $H_1$ is the union over $n_0$ rows of all $m$-tuples of nodes within each row. The same with 
columns and diagonals. 
Explicitly, we define two functions $u,v$ that associate to  any $i\in \{1,\dots,n_0\}$ its coordinates $(u(i),v(i))$ on
the $n_0\times n_0$ grid. Then, the $H_j$ are $m$-tuple of nodes defined as 
\begin{align*}
  H_1 &=\cup_{u=1}^{n_0}H_1(u)=\cup_{u=1}^{n_0}\{ \{i_1,\dots,i_m\} \in \V^{(m)}~|~ \forall k, u(i_k)=u, v(i_k)\in\{1,\cdots,n_0\}\},\\
  H_2 &=\cup_{v=1}^{n_0}H_2(v)=\cup_{v=1}^{n_0}\{ \{i_1,\dots,i_m\} \in \V^{(m)}~|~ \forall k, v(i_k)=v, u(i_k)\in\{1,\cdots,n_0\}\},\\
  H_3 &=\cup_{s=1}^{n_0}H_3(s)\\
  &=\cup_{s=1}^{n_0}\{ \{i_1,\dots,i_m\} \in \V^{(m)}~|~ \forall k, u(i_k)=s, v(i_k)= s+t  \operatorname{mod} n_0
          \text{ for some }      t   \in\{1,\cdots,n_0\}\}. 
\end{align*}
%%%%
The $H_j$ are pairwise disjoints as required. 

The matrix $M_j$ of conditional probabilities of $U_j$ given $Z$
 has $Q^n$ rows indexed by composite states of all $n=n_0^2$ nodes, and
$2^{n_0 \binom{n_0}{m}}$ columns indexed by $m$-tuples in $H_j$.

Observe that with an appropriate ordering of the rows and columns (which is
dependent on $j$), $M_j$ has a block structure given by
\begin{equation}
  M_j=A\otimes A \otimes \cdots \otimes A \text{ ($n_0$
    factors)}.\label{eq:B1}
\end{equation}
(Note that since $A$ is
$Q^{n_0}\times 2^{\binom{n_0}{m}}$, the tensor product on the right is
$(Q^{n_0})^{n_0}\times \left (2^ {\binom{n_0}{m}} \right )^{n_0}$ which is
$Q^{n_0^2}\times 2^ {n_0 \binom{n_0}{m}}$, the size of $M_j$.)
That $M_j$ is this tensor product is most easily seen by noting the partitioning
of the $n_0^2$ nodes into $n_0$ disjoint sets (rows, columns and diagonals of the grid)  gives rise to $n_0$ copies of
the matrix $A$, one for each set of all simple $m$-uniform hypergraphs over $n_0$ nodes.  The row indices of $M_j$ are obtained by choosing an assignment of
states to the nodes in $H_j(u)$ for each $u$ independently, and the
column indices by the union of independently-chosen simple $m$-uniform hypergraphs subgraphs on $H_j(u)$ for each $u$. This independence in both
rows and columns leads to the tensor decomposition of $M_j$.

Now since $A$ has generically full row rank ($Q^{n_0}$), equation (\ref{eq:B1}) implies that $M_j$ does as well
(\emph{i.e} has row rank $Q^{n_0^2}=Q^n$).
\end{proof}

Next, with $\mathbf{v},M_1,M_2,M_3$ defined by the embedding given in the previous paragraphs, we apply Kruskal's Theorem to the table $[\mathbf{v};M_1,M_2,M_3]$. By construction of the $M_j$, condition
\eqref{eq:kruskal_condition} is generically satisfied since $3Q^n \geq 2Q^n +2$.  Thus
the vector $\mathbf{v}$ and the matrices $M_1,M_2,M_3$ are generically uniquely determined, up to simultaneous permutation of the
rows from the distribution of a simple $m$-uniform HSBM.

It now remains  to recover the original parameters
of the simple $m$-uniform HSBM: 
the group proportions $\pi_q$ and the connectivity matrix $(\Bqm)_{1\le q_1\le q_m\le Q}$. %Due to the specific form of the parameters, $(\Bqm)_{1\le q_1\le q_m\le Q}$, this part will slightly differ from the one in the proof of Theorem 2 in \cite{allm:etal:11}.

\paragraph*{Conclusion for the original model.}
%%%%%%%%%%%
The entries of $\mathbf{v}$ are of the form
$\pi_1^{n_1}\cdots\pi_Q^{n_Q}$ with $\sum n_q =n$, while the entries of the $M_j$ contain information on the $\Bqm$. Although
the ordering of the rows of the $M_j$ is arbitrary, crucially we do know how the rows of $M_j$ are paired with the entries of $\mathbf v$.

By focusing on one of the matrices, say  $M_1$, and adding appropriate columns of it, we can obtain marginal conditional
probabilities of single hyperedge variables, namely a column vector with values $(\Pt(\Ym=1| (Z_1,\dots,Z_n)=v))_v$ for
any $m$-tuple $\{i_1,\dots, i_m\}$. Indeed, this vector is obtained by summing all the columns of $M_1$ corresponding to simple $m$-uniform hypergraphs with
$\Ym=1$. Thus, we recover the 
set of values $\{\Bqm\}_{1\le q_1 \le \dots \le q_m \le Q}$, but without order. Namely, we still do not know the $\Bqm$
up to a permutation on $\{1,\dots,Q\}$ only, but rather up to a permutation on $\{1,\dots,Q\}^n$. 

In the following, we assume without loss of generality, as it is a generic condition, that all $\{\Bqm\}_{1\le q_1 \le  \dots \le q_m \le Q}$ are distinct. 

We look at the first $(m+1)$ nodes $\V_1= \{1,\dots, m, m+1\}$ and consider the $m+1$ different $m$-tuples $\{i_1,\dots,
i_m\}\in \V_1^{(m)}$ that can be made from  these nodes ($i_k\in \V_1$).
Again, for each of these $m$-tuples, adding appropriate columns of
$M_1$, we can jointly obtain  the vectors of conditional marginal probabilities
$(\Pt(Y_{\{i_1,\dots, i_{m}\}}=1 | (Z_1,\dots,Z_n)=v))_v$. Jointly means that all those vectors  share the same ordering over the index $v\in \{1,\dots,Q\}^n$. 
In other words, we recover the sets of values 
\[
\forall v\in \{1,\dots, Q\}^n, \quad   R_v =  \{B^{(m,n)}_{v_{i_1},\dots,v_{i_{m}}} ; \{i_1,\dots,
i_m\} \in \V_1^{(m)}\}.
\]
Now, we assumed the $B$'s are all distinct so the cardinalities of the sets $R_v$ will help us discriminate the
different parameters (up to a permutation on $\{1,\dots, Q\}$ only). Indeed, there are exactly $Q$ sets $R_v$ with
cardinality exactly one. These correspond to the cases were $v=(q,q,\dots,q)$ for some $1\le q\le Q$.
From this, we can distinguish the parameters of the form
$\{B^{(m,n)}_{q,\dots,q} ; 1\le q\le Q \}$ from the complete set of parameters. Note that the corresponding entries of $v$
are given by $\pi_q^m$. So we also recover the paired values $\{(\pi_q,B^{(m,n)}_{q,\dots,q}) ; 1\le q\le Q \}$. 
Then, we continue with the sets $R_v$ with cardinality two: these  are of the form $\{B^{(m)}_{q,\dots,q} ; B^{(m,n)}_{q,\dots,q,l}\}$ for some
$1\le q\neq l\le Q$. As we already identified the parameters $\{B^{(m,n)}_{q,\dots,q} ; 1\le q\le Q \}$ and all $B$'s are
distinct, this enables us to identify the set of parameters $\{B^{(m,n)}_{q,\dots,q,l} ; 1\le q\neq l\le Q\}$. By induction,
we recover the set of parameters $\{B^{(m,n)}_{q,\dots,q,l,l'} ; 1\le q,l,l'\le Q \text{ and } q,l,l' \text{ distinct}\}$ \emph{et caetera}, ending with the set of
parameters $\{B^{(m,n)}_{q_1,\dots,q_m} ; 1\le q_1<q_2<\dots <q_m\le Q\}$. This means that we finally have obtained the
parameters  $\{\pi_q,\Bqm\}_{1\le q\le Q ; 1\le q_1 \le \dots \le q_m \le Q}$ up to a permutation over $\{1,\dots, Q\}$.

Finally, note that all generic aspects of this argument, in the base case and the requirement that the parameters $\Bqm$ be distinct, concern only the $\Bqm$. Thus if the group proportions $\pi_q$ are fixed to any specific values, the theorem remains valid. 

%\begin{rmk}
  The requirement on large enough $n$ is more precisely given as $n\geq Q^2p^2$ where $p$ is the smallest integer such
  that $\binom{p-1}{m-1}\geq Qm$. 
 \blue{By choosing 
\[
(p-m+1)^{m-1} \ge m! Qm,
\]
we ensure that the condition on $p$ is satisfied. This ensures identifiability  as %LB soon 
long as 
\[
n \ge Q^2 \Big(m! Qm +m -1\Big)^{2/(m-1)}.
\]
}
  %A rough approximation gives that $p$ is of the order $(Qm)^{1/(m-1)}$ which gives that $n$ should be larger than $Q^2(Qm)^{2/(m-1)}$. 
%\end{rmk}

%%%%%%%%%%%%%
%%%%%%%%%%%%%
\section{Other proofs}
%%%%%%%%%%
\label{app:proofs}

\begin{proof}[Proof of Lemma~\ref{lem:nb}]
We consider a fixed value of $m\ge 2$ and denote by $ \llbracket a, b\rrbracket$ the set of integer values between $a,b$. 
Let us recall that $B$ is a fully symmetric tensor \eqref{eq:sym_tensor}, so the number of free parameters in $B$ is equal to the number of ordered sequences $q_1\le \dots\le q_m$ of elements in $\llbracket 1, Q\rrbracket$. We denote by $\mathcal{Q}^+$ this set. Then we define a function $f$ which, to any such sequence $\underline q = (q_1,\dots, q_m)$, associates the value $\underline l= f(\underline q)$ defined by  $f(\underline q)=(q_1,q_2+1,q_3+2,\dots, q_m+m-1)$.  We let $\mathcal{L}^+$ denote the set of sequences $\underline l=(l_1,\dots, l_m)$ with coordinates in $\llbracket 1, Q+m-1\rrbracket$ and such that $l_1<l_2<\dots <l_m$.
Thus, for any $\underline q \in \mathcal{Q}^+$ we get that $f(\underline q)\in \mathcal{L}^+$.

Conversely, for any $\underline l =(l_1,\dots,l_m)\in \mathcal{L}^+$, we can associate the value $\underline q = g(\underline l)= (l_1, l_2-1,l_3-2, \dots, l_m-m+1)$. It is easy to see that the image $\underline q= g(\underline l)$ belongs to $\mathcal{Q}^+$. 

As a consequence, the functions $f$ and $g$ are such that their composition is the identity function: $f \circ g=g\circ f = Id$. These are one-to-one functions mapping $\mathcal{Q}^+$ to $\mathcal{L}^+$ and conversely. This implies that the cardinalities of these two sets are equal. But an element in $\mathcal{L}^+$ is exactly a subset of size $m$ of $\llbracket 1, Q+m-1\rrbracket$ so that 
the cardinality of $\mathcal{L}^+$ is the number of subsets of  size $m$ of $\llbracket 1, Q+m-1\rrbracket$. This concludes the proof of the lemma. 
\end{proof}

%%%%%%%%
%%%%%%%%

\begin{proof}[Proof of Corollary~\ref{cor:ident}]
  From the probability distribution $\Pt$ over simple hypergraphs $\mathcal{H}$ on a set of $n$ nodes and hyperedges
  with largest size $M$, we automatically obtain all the probability distributions $\Pt$ restricted to  simple $m$-uniform
  hypergraphs $\mathcal{H}_m$ on the same set of nodes. Applying the result of Theorem~\ref{thm:ident} for all values
  $m$ is sufficient to obtain the desired result. Indeed, as $M$ is finite, the union of the finite number of  lower-dimensional 
  subspaces where identifiability for fixed $m$ may not be satisfied gives a lower-dimensional subspace, ensuring
  generic identifiability. Moreover, for each value of $m$, we recover the parameter $\theta^{(m)}$ up to a permutation
  on $\{1,\dots, Q\}$. Now, for any $m\neq m'$ it remains to be able to jointly order the parameters $\theta^{(m)}$ and
  $\theta^{(m')}$ up to a permutation on $\{1,\dots,Q\}$. 
If all the $\pi_q$'s are different, which is a generic condition, this can be easily done because $\theta^{(m)}$ and $\theta^{(m')}$ share the same
distinct $\pi_q$'s. 
\end{proof}

\begin{proof}[Proof of Proposition~\ref{prop:VEstep}]
We want to maximize $ \mathcal{J}({\theta}, {\tau}) $ with respect to $ \tau_{iq} $ under the constraint $ \sum_{q=1}^{Q}\tau_{iq} = 1 $ for all $ i $. Using the method of Lagrange multipliers, this is equivalent to maximizing with respect to $\tau_{iq}$ the Lagrangian function
\begin{align*}
\Lambda({\theta}, {\tau}, {\lambda}) 
&= \sum_{i=1}^{n} \lambda_i \Bigg( \sum_{q=1}^{Q} \tau_{iq}-1 \Bigg) + \mathcal{J}({\theta}, {\tau}) \\
&= \sum_{i=1}^{n} \lambda_i \Bigg( \sum_{q=1}^{Q} \tau_{iq}-1 \Bigg) + \sum_{q=1}^{Q}\sum_{i=1}^{n}\tau_{iq}\log\frac{\pi_q}{\tau_{iq}} \\
&\phantom{=} + \sum_{i=1}^{n}\sum_{q=1}^{Q}\sum_{m=1}^{M-1}\sum_{q_1\le \dots\le q_m }\sum_{\Vm\not\ni i} \tau_{iq}\tau_{i_1q_1} \cdots \tau_{i_mq_m} 
f(Y_{ii_1 \dots i_m}, B^{(m+1,n)}_{qq_1 \dots q_m})
%\left[ Y_{ii_1 \dots i_m}\log B^{(m+1)}_{qq_1 \dots q_m} \right.\\
%&\qquad \qquad \qquad \left. + (1-Y_{ii_1 \dots i_m})\log(1-B^{(m+1)}_{qq_1 \dots q_m}) \right].
\end{align*}
		Computing the partial derivative of $ \Lambda(\theta, \bm{\tau}, \bm{\lambda}) $ with respect to $ \tau_{iq} $, we obtain the following expression
\begin{align*}
\frac{\partial\Lambda}{\partial\tau_{iq}} 
	&= \lambda_i + \log\frac{\pi_q}{\tau_{iq}} - 1 \\
	&\phantom{=} + \sum_{m=1}^{M-1}\sum_{q_1\le \dots\le q_m}\sum_{\V^m \not\ni i} \tau_{i_1q_1} \cdots \tau_{i_mq_m} 
	f(Y_{ii_1 \dots i_m}, B^{(m+1,n)}_{qq_1 \dots q_m})\\
	&= \lambda_i + \log\pi_q - \log\tau_{iq} - 1 \\
	&\phantom{=} + \log\prod_{m=1}^{M-1}\prod_{q_1\le \dots\le q_m}\prod_{\V^m \not\ni i} \left[ \exp( f(Y_{ii_1 \dots i_m}, B^{(m+1,n)}_{qq_1 \dots q_m}))
	%(B^{(m)}_{qq_1 \dots q_m})^{Y_{ii_1 \dots i_m}} \cdot (1-B^{(m)}_{qq_1 \dots q_m})^{1-Y_{ii_1 \dots i_m}} 
	\right]^{\tau_{i_1q_1} \cdots \tau_{i_mq_m}},
		\end{align*}
		which is equal to 0 if
\begin{equation*}
			\tau_{iq} = e^{\lambda_i - 1}\ \pi_q \prod_{m=1}^{M-1}\prod_{q_1\le \dots\le q_m}\prod_{\V^m \not\ni i} \left[ (\exp( f(Y_{ii_1 \dots i_m}, B^{(m+1,n)}_{qq_1 \dots q_m})) \right]^{\tau_{i_1q_1} \cdots \tau_{i_mq_m}}.
		\end{equation*}
The term $ e^{\lambda_i-1} = \frac{1}{\sum_{q=1}^{Q}\tau_{iq}} $ is the normalizing constant such that $ \sum_{q=1}^{Q}\tau_{iq} = 1 $ for each $ i $.\\
Finally, let us remark that the Lagrangian function $\Lambda$ is concave with respect to each $\tau_{iq}$, being the sum of a concave term ($ \tau_{iq} \log(\pi_q / \tau_{iq}) $) and linear terms. Thus the critical point is a maximum.
\end{proof}

%%%%
\begin{proof}[Proof of Proposition~\ref{prop:Mstep}]
For the prior probabilities $ \pi_q $, we want to maximize $ \mathcal{J}({\theta}, {\tau}) $ with respect to $ \pi_q $ subject to the constraint $ \sum_{q=1}^{Q}\pi_q = 1 $. Using again Lagrange multipliers, this is equivalent to maximizing 
\begin{equation*}
\Lambda({\theta}, {\tau}, {\lambda}) = \lambda \Bigg( \sum_{q=1}^{Q}\pi_q - 1 \Bigg) + \mathcal{J}({\theta}, {\tau})
\end{equation*}
Noting that the second term of $ \mathcal{J}({\theta}, {\tau}) $ does not depend on $ \pi_q $, the computation of the partial derivative of $ \Lambda({\theta}, {\tau}, {\lambda}) $ reduces to
\begin{equation*}
\frac{\partial}{\partial\pi_q} \Bigg[ \lambda \bigg( \sum_{q=1}^{Q}\pi_q - 1 \bigg) + \sum_{q=1}^{Q}\sum_{i=1}^{n}\tau_{iq}\log\frac{\pi_q}{\tau_{iq}} \Bigg] = \lambda + \sum_{i=1}^{n} \frac{\tau_{iq}}{\pi_q}.
\end{equation*}
This quantity results equal to 0 if
\begin{equation*}
\pi_q = -\frac{1}{\lambda} \sum_{i=1}^{n}\tau_{iq},
\end{equation*}
where $ \lambda = - n $ is the normalizing constant in order to satisfy $ \sum_{q=1}^{Q}\pi_q = 1 $.\\
Note the Lagrangian function $\Lambda$ is concave with respect to each $\pi_q$, being the sum of a concave term ($\log(\pi_q / \tau_{iq})$), of a linear term ($\lambda\sum_{q=1}^{Q}\pi_q$) and of a constant. The critical point is then a maximum.
		
Finally, the partial derivative \emph{w.r.t.} $\Bqm$ is 
\begin{equation*}
\frac{\partial\mathcal{J}}{\partial \Bqm} = \sum_{\V^m} \tau_{i_1q_1} \cdots \tau_{i_mq_m} \left[ Y_{i_1 \dots i_m}\frac{1}{B_{q_1 \dots q_m}} - (1-Y_{i_1 \dots i_m})\frac{1}{1-B_{q_1 \dots q_m}} \right].
\end{equation*}
Through some basic algebraic manipulations, this quantity results equal to 0 if
\begin{equation*}
\Bqm = \frac{\sum_{\V^m} \tau_{i_1q_1} \cdots \tau_{i_mq_m} Y_{i_1 \dots i_m}}{\sum_{\V^m} \tau_{i_1q_1} \cdots \tau_{i_mq_m}}.
\end{equation*}
Again, the Lagrangian function is the sum of a concave term ($\log \Bqm$) and of some constant terms, thus being a concave function. The critical point is then a maximum.
\end{proof}

\begin{proof}[Proof of Proposition~\ref{prop:Mstep-aff}] 
The following decomposition of $ \mathcal{J}(\theta, \tau) $ naturally holds:
		\begin{align*}
			\mathcal{J}(\theta, \tau)
			&= \sum_{q=1}^{Q}\sum_{i=1}^{n}\tau_{iq}\log\frac{\pi_q}{\tau_{iq}} \\
			&\phantom{=} + \sum_{m=2}^{M}\sum_{q=1}^{Q}\sum_{\V^m} \tau_{i_1q} \cdots \tau_{i_mq} \left[ \Ym\log \alpha^{(m)} + (1-\Ym)\log(1-\alpha^{(m)}) \right] \\
			&\phantom{=} + \sum_{m=2}^{M}\sum_{\substack{q_1\le \dots \le q_m \\|\{q_1,\dots,q_m\}|\ge 2}}\sum_{\V^m} \tau_{i_1q_1} \cdots \tau_{i_mq_m} \left[ \Ym\log \beta^{(m)} + (1-\Ym)\log(1-\beta^{(m)}) \right].
		\end{align*}
		The partial derivative \emph{w.r.t.} $\alpha^{(m)}$ is 
		\begin{equation*}
			\frac{\partial\mathcal{J}}{\partial\alpha^{(m)}} = \sum_{q=1}^{Q}\sum_{\V^m} \tau_{i_1q} \cdots \tau_{i_mq} \left[ Y_{i_1 \dots i_m}\frac{1}{\alpha^{(m)}} - (1-Y_{i_1 \dots i_m})\frac{1}{1-\alpha^{(m)}} \right],
		\end{equation*}
		hence it follows that:
		\begin{equation*}
			\widehat {\alpha}^{(m)} = \frac{\sum_{q=1}^Q\sum_{\{i_1,\dots,i_m\}\in \V^m} \tau_{i_1q} \dots \tau_{i_mq} Y_{i_1 \dots i_m}}{\sum_{q=1}^Q\sum_{\{i_1,\dots,i_m\}\in \V^m} \tau_{i_1q} \dots \tau_{i_mq}}.
		\end{equation*}
		Analogously, the partial derivative \emph{w.r.t.} $\beta^{(m)}$ is 
		\begin{equation*}
			\frac{\partial\mathcal{J}}{\partial\beta^{(m)}} = \sum_{\substack{q_1\le \dots \le q_m \\|\{q_1,\dots,q_m\}|\ge 2}}\sum_{\V^m} \tau_{i_1q_1} \cdots \tau_{i_mq_m} \left[ Y_{i_1 \dots i_m}\frac{1}{\beta^{(m)}} - (1-Y_{i_1 \dots i_m})\frac{1}{1-\beta^{(m)}} \right],
		\end{equation*}
		and
		\begin{equation*}
			\widehat {\beta}^{(m)} = \frac{\sum_{\substack{q_1\le \dots \le q_m \\|\{q_1,\dots,q_m\}|\ge 2}}\sum_{\{i_1,\dots,i_m\}\in \V^m} \tau_{i_1q_1} \dots \tau_{i_mq_m} Y_{i_1 \dots i_m}}{\sum_{\substack{q_1\le \dots \le q_m \\|\{q_1,\dots,q_m\}|\ge 2}}\sum_{\{i_1,\dots,i_m\}\in \V^m} \tau_{i_1q_1} \dots \tau_{i_mq_m}}.
		\end{equation*}
		This concludes the proof for the formulas under assumption \eqref{eq:aff-m}. 
		The expressions for $ \hat \alpha $ and $ \hat \beta $ under assumption \eqref{eq:aff} are computed in the same way. 
\end{proof}

%%%%%%%%%%%%%
%%%%%%%%%%%%%
\section{Computational details on the algorithm's implementation}
\label{app:details}
In order to provide an efficient implementation, the whole estimation algorithm is implemented in {\bf\CC} language using the \texttt{Armadillo} library for linear algebra. Moreover the implementation is made available in \texttt{R} by means of the \texttt{R} packages \texttt{Rcpp} \citep{Rcpp1, Rcpp2} and \texttt{RcppArmadillo} \citep{RcppArmadillo}. In the following we consider some of the most relevant computational details. 

\paragraph{Dealing with heavy computational cost.} 
Dealing with very large data structures, the main drawback of the proposed algorithm is the intensive computational effort, in terms of both execution time needed to converge and required memory space. The most outstanding example regards the computation of the products $\tau_{i_1q_1} \cdots \tau_{i_mq_m}$, required both in the \texttt{VE}-Step (see Proposition \ref{prop:VEstep}, for $\tau_{iq}$) and in the \texttt{M}-Step (see Proposition \ref{prop:Mstep}, for $\Bqm$). The huge computational cost of this calculation derives from the large number of potential unordered node tuples even for rather small values of $n$ and $m$; indeed 
$| \mathcal{V}^{(m)} | = \binom{n}{m}$.
A first possibility is to compute all the products $\tau_{i_1q_1} \cdots \tau_{i_m q_m}$ in a recursive manner at the beginning of each \texttt{VEM} iteration and to store them in a matrix. Although this is actually very beneficial for the computational time, the resulting matrix is huge, having  number of rows and columns equal to $\binom{n}{m}$ \blue{and $Q^m$ respectively}. 
% Below this is true but inefficient: 
%Another possibility would be  to store these values in a matrix with $n!/(n-m)!$ rows and $\binom{Q+m-1}{m}$ columns, respectively. None of these solutions are tractable,  except for very small values of $n$, $Q$, and (especially) $m$.}
%%%%%%%
%\red{having  number of rows and columns equal to $\binom{n}{m}$ and  $\binom{Q+m-1}{m}$ NO, should be  $Q^m$ for this number of rows. Or the other way round would be to store these values in a matrix with $n!/(n-m)!$ rows and $\binom{Q+m-1}{m}$ columns} respectively. The result is a structure that is intractable except for very small values of $n$, $Q$, and (especially) $m$. 
Taking into account that every element requires 8 bytes, we report some examples in Table \ref{tab:size_prod}, in order to better clarify the magnitude of the quantity to store. 
Stated the impossibility to store a matrix of such size, the computation of the required products $\tau_{i_1q_1} \cdots \tau_{i_mq_m}$ is implemented directly inside the \texttt{VE}- and \texttt{M}-Steps through nested loops; this process involves an important increase in the computing times, but on the other hand requires a minimal amount of memory. To handle the slowness of the computation, both the \texttt{VE}-Step and the \texttt{M}-Step are efficiently implemented in parallel through the \texttt{RcppParallel} package \citep{RcppParallel}.
\bigskip
\begin{table}[htbp]
\caption{Memory size of the matrix containing the products $\tau_{i_1q_1} \cdots \tau_{i_m q_m}$ for given values of $n$ (number of nodes), $Q$ (number of latent groups) and $m$ (hyperedge size).}
\centering
\begin{tabular}{C{1cm}C{1cm}C{1cm}C{2.5cm}C{1cm}C{1cm}C{1cm}C{1cm}C{2.5cm}}
	\cmidrule(lr){1-4}\cmidrule(lr){6-9}
	$ n $   & $ m $ & $ Q $ & Memory size               & & $ n $ & $ m $ & $ Q $ & Memory size            \\
	\cmidrule(lr){1-4}\cmidrule(lr){6-9}
	50      & 3     & 2		& $\approx 1.25\ MB$        & & 150   & 3     & 2     & $\approx 35.28\ MB$    \\
	50		& 3 	& 3  	& $\approx 4.23\ MB$        & & 150   & 3     & 3     & $\approx 119.08\ MB$   \\
	100 	& 3 	& 2  	& $\approx 10.34\ MB$       & & 200   & 3     & 2     & $\approx 84.05\ MB$    \\
	100		& 3 	& 3  	& $\approx 34.93\ MB$       & & 200   & 3     & 3     & $\approx 283.69\ MB$   \\
	\cmidrule(lr){1-4}\cmidrule(lr){6-9}
\end{tabular}
\label{tab:size_prod}
\end{table}
\bigskip

\paragraph{Floating point underflow.} Another crucial aspect is the possible occurrence of numerical instability deriving from the multiplication of many small values in the computation of $\widehat{\tau}_{iq}$. %As suggested in Proposition \ref{prop:VEstep}, 
A simple remedy is provided by the calculation of $\log\widehat{\tau}_{iq}$ instead of $\widehat{\tau}_{iq}$.  
%The final solution, denoted by $\widetilde{\tau}_{iq} = \log\widehat{\tau}_{iq}$, relies on the following formula:
%
%\begin{equation*}
%    \widehat{\tau}_{iq} = \frac{\exp(\widetilde{\tau}_{iq} - \widetilde{\tau}_{\textsc{max}, i})}{\sum_{p=1}^{Q}\exp(\widetilde{\tau}_{iq} - \widetilde{\tau}_{\textsc{max}, i})},
%\end{equation*}
So, denoting $b_{iq}=\log(\hat \tau_{iq})-c_i$, we compute $\hat \tau_{iq}$ relying on 
\begin{equation*}
    \widehat{\tau}_{iq} = \frac{\exp(b_{iq} - b_{\textsc{max}, i})}{\sum_{p=1}^{Q}\exp(b_{iq} - b_{\textsc{max}, i})},
\end{equation*}
where $b_{\textsc{max}, i} = \underset{q=1 \ldots Q}{\max} b_{iq}$ prevents the denominator to grow excessively large, thus avoiding new potential numerical issues related to the floating point underflow.

\section{HSBM simulations details and additional analyses}
\subsection{Settings}
\label{sec_SM:synth_pram}
We generated hypergraphs under the HSBM  with two or three latent groups ($Q = 2$ or $Q = 3$). The group proportions were non-uniform, with $\pi = (0.6, 0.4)$ for $Q = 2$ and $\pi = (0.4, 0.3, 0.3)$ for $Q = 3$. We set the largest hyperedge size $M$ to 3, and we considered different numbers of nodes, $n \in \{50, 100, 150, 200\}$.
To simplify the latent structure, we assumed the \eqref{eq:aff-m} submodel, and we parameterized the model through the ratios $\rho^{(m)}$ of within-group size-$m$ hyperedges over between-groups size-$m$ hyperedges. More precisely,	we let $E^{(m,n)}_{\text{within}}$ (resp. $E^{(m,n)}_{\text{between}}$) denote the number of within-group (resp. between-groups) size-$m$ hyperedges. We have that
\begin{equation*}
    E^{(m,n)}_{\text{within}} = \binom{n}{m} \alpha^{(m,n)}\sum_q \pi_q^m \quad \text{ and } \quad
    E^{(m,n)}_{\text{between}} = \binom{n}{m} \beta^{(m,n)}(1-\sum_q \pi_q^m) 
\end{equation*}
\blue{where the notation $\alpha^{(m,n)}$ (resp. $\beta^{(m,n)}$) underlines the dependency of the parameter $\alpha^{(m)}$ (resp. $\beta^{(m)}$) to the sample size $n$.} We also obtain the ratio 
\[
\rho^{(m,n)}=\frac{E^{(m,n)}_{\text{within}}}{E^{(m,n)}_{\text{between}}} = \frac{\alpha^{(m,n)}\sum_q \pi_q^m } {\beta^{(m,n)}(1-\sum_q \pi_q^m) }.
\]
To simulate sparse hypergraphs, we choose the values $\alpha^{(2,n)}, \beta^{(2,n)}$ (resp. $\alpha^{(3,n)}, \beta^{(3,n)}$) to be constant divided by $n$	(resp. constant divided by $n^2$). This implies that the numbers of within and between-groups  size-$m$ hyperedges grow linearly with $n$.
We thus constrain the parameters with the following forms:
\begin{align*}
\alpha^{(2,n)}&=\alpha_0 \times 50/n , \qquad \alpha^{(3,n)}=c \alpha_0 \times50/n^2 ,\\
\beta^{(2,n)}&=\beta_0 \times 50/n , \qquad \beta^{(3,n)}=\beta_0 \times50/n^2 ,
\end{align*}
for some constant $c$ to be specified. 
With these choices, we get that the ratio $\rho^{(m,n)}$ does not depend on $n$ and equals:
\[
\rho^{(m)}= c^{m-2}\frac{\alpha_0 \sum_q \pi_q^m }{\beta^0(1-\sum_q \pi_q^m)}.
\]
To choose the parameter values of our scenarios, we thus start by setting $\alpha_0$ and a constant ratio $\rho$ that will be used both for pairwise and three-wise interactions ($m=2,3$). Then we simply obtain 
\[
    \beta_0 = \frac {\alpha_0}{\rho}\frac {\sum_q \pi_q^2}{1-\sum_q \pi_q^2}\quad \text{and} \quad
    c= \frac {\sum_q \pi_q^2}{1-\sum_q \pi_q^2} \times \frac {1-\sum_q \pi_q^3}{\sum_q \pi_q^3}.
\]
	
The initial choices of $\alpha_0, \rho$ in each setting are reported in Table~\ref{tab:sim_setting}. The resulting parameter values and number of expected hyperedges are reported in Tables~\ref{tab:sim_param_A} and ~\ref{tab:sim_param_B}.	Note that while settings A2 and A3' have same $\alpha_0,\rho$ parameter values, they differ through the number of groups $Q$.

% Descrition of settings
%\begin{table}[htbp]
%    \centering
%    \begin{tabular}{C{1cm}C{2cm}C{2cm}C{2cm}C{2cm}C{3cm}}
%    \toprule
%         & \multicolumn{2}{c}{Scenarios A} & \multicolumn{2}{c}{Scenarios B} & Scenario A3' \\
%         & $Q=2$& $Q=3$ & $Q=2$ & $Q=3$ & $Q=3$ \\
%         \cmidrule(r){2-3}\cmidrule(r){4-5}\cmidrule(r){6-6}
%         $\alpha_0$ & 0.70 & 0.30 & 0.30 & 0.40 & 0.70\\
%         $\rho$     & 1.20 & 1.70 & 0.50 & 0.25& 1.20 \\
%	\bottomrule         
%    \end{tabular}
%    \caption{Description of the simulated scenarios.}
%    \label{tab:sim_setting}
%\end{table}

% Descrition of settings
\begin{table}[htbp]
\caption{Description of the simulated scenarios: scenarios A exhibit community structure ($\rho >1$) while scenarios B exhibit disassortative behaviours ($\rho<1$). The numbering of the settings (e.g ``2'' in  A2) indicates the number of groups $Q$.}
    \centering
    \begin{tabular}{C{1cm}C{2cm}C{2cm}C{2cm}C{2cm}C{2cm}}
    \toprule
         & \multicolumn{3}{c}{Scenarios A} & \multicolumn{2}{c}{Scenarios B}  \\
         \cmidrule(r){2-4}\cmidrule(r){5-6}
         & A2& A3 & A3' & B2 & B3 \\
         & ($Q=2$) & ($Q=3$) & ($Q=3$) & ($Q=2$) & ($Q=3$)  \\
         \cmidrule(r){2-4}\cmidrule(r){5-6}
         $\alpha_0$ & 0.70 & 0.30 & 0.70 & 0.30 & 0.40 \\
         $\rho$     & 1.20 & 1.70 & 1.20 & 0.50 & 0.25 \\
	\bottomrule         
    \end{tabular}
    \label{tab:sim_setting}
\end{table}

% Scenario A
\begin{table}[htbp]
\caption{Resulting parameter values (within-group $\alpha^{(m,n)}$ and between-groups $\beta^{(m,n)}$, rounded at the fifth decimal) and expected number $E^{(m,n)}$ of size-$m$ hyperedges (rounded at integer value) for Scenarios A and either $Q=2$ (setting A2) or   $Q=3$ (settings A3 and A3').}
\centering
\begin{tabular}{C{3cm}C{2cm}C{2cm}C{2cm}C{2cm}}
    \toprule
    Setting A2          & $n=50$    & $n=100$   & $n=150$   & $n=200$   \\
    \toprule
    $\alpha^{(2,n)}$    & 0.70000   & 0.35000   & 0.23333   & 0.17500   \\
    $\beta^{(2,n)}$     & 0.63194   & 0.31597   & 0.21065   & 0.15799   \\
    $\alpha^{(3,n)}$    & 0.03900   & 0.00975   & 0.00433   & 0.00244   \\
    $\beta^{(3,n)}$     & 0.01264   & 0.00316   & 0.00140   & 0.00079   \\
    \midrule
    $E^{(2,n)}$         & 817       & 1652      & 2486      & 3320      \\
    $E^{(3,n)}$         & 392       & 809       & 1226      & 1643      \\
    \bottomrule
    \toprule
    Setting A3          & $n=50$    & $n=100$   & $n=150$   & $n=200$   \\
    \toprule
    $\alpha^{(2,n)}$    & 0.30000   & 0.15000   & 0.10000   & 0.07500   \\ 
    $\beta^{(2,n)}$     & 0.09091   & 0.04545   & 0.03030   & 0.02273   \\ 
    $\alpha^{(3,n)}$    & 0.02310   & 0.00578   & 0.00257   & 0.00144   \\ 
    $\beta^{(3,n)}$     & 0.00182   & 0.00045   & 0.00020   & 0.00011   \\
    \midrule
    $E^{(2,n)}$         & 198       & 401       & 603       & 806       \\ 
    $E^{(3,n)}$         & 85        & 175       & 265       & 355       \\
	\bottomrule
	 \toprule
	    Setting A3'         & $n=50$    & $n=100$   & $n=150$   & $n=200$   \\
    \toprule
    $\alpha^{(2,n)}$    & 0.70000   & 0.35000   & 0.23333   & 0.17500   \\ 
    $\beta^{(2,n)}$     & 0.30051   & 0.15025   & 0.10017   & 0.07513   \\ 
    $\alpha^{(3,n)}$    & 0.05391   & 0.01348   & 0.00599   & 0.00337   \\ 
    $\beta^{(3,n)}$     & 0.00601   & 0.00150   & 0.00067   & 0.00038   \\ 
    \midrule
    $E^{(2,n)}$         & 535       & 1080      & 1625      & 2171      \\ 
    $E^{(3,n)}$         & 229       & 471       & 714       & 957       \\ 
	\bottomrule
\end{tabular}
\label{tab:sim_param_A}
\end{table}
%

% Scenario B
\begin{table}[htbp]
\caption{Resulting parameter values (within-group $\alpha^{(m,n)}$ and between-groups $\beta^{(m,n)}$, rounded at the fifth decimal) and expected number $E^{(m,n)}$ of size-$m$ hyperedges (rounded at integer value) for Scenarios B and either $Q=2$ (setting B2) or   $Q=3$ (setting B3).}
\centering
\begin{tabular}{C{3cm}C{2cm}C{2cm}C{2cm}C{2cm}}
	\toprule
    Setting B2          & $n=50$    & $n=100$   & $n=150$   & $n=200$   \\
    \toprule
    $\alpha^{(2,n)}$    & 0.30000   & 0.15000   & 0.10000   & 0.07500   \\
    $\beta^{(2,n)}$     & 0.65000   & 0.32500   & 0.21667   & 0.16250   \\
    $\alpha^{(3,n)}$    & 0.01671   & 0.00418   & 0.00186   & 0.00104   \\
    $\beta^{(3,n)}$     & 0.01300   & 0.00325   & 0.00144   & 0.00081   \\
    \midrule
    $E^{(2,n)}$         & 573       & 1158      & 1743      & 2328      \\
    $E^{(3,n)}$         & 275       & 568       & 860       & 1153      \\
    \bottomrule
    \toprule
    Setting B3          & $n=50$    & $n=100$   & $n=150$   & $n=200$   \\
    \toprule
    $\alpha^{(2,n)}$    & 0.40000   & 0.20000   & 0.13333   & 0.10000   \\ 
    $\beta^{(2,n)}$     & 0.82424   & 0.41212   & 0.27475   & 0.20606   \\ 
    $\alpha^{(3,n)}$    & 0.03080   & 0.00770   & 0.00342   & 0.00193   \\ 
    $\beta^{(3,n)}$     & 0.01648   & 0.00412   & 0.00183   & 0.00103   \\ 
    \midrule
    $E^{(2,n)}$         & 833       & 1683      & 2533      & 3383      \\ 
    $E^{(3,n)}$         & 356       & 735       & 1113      & 1492      \\
	\bottomrule
\end{tabular}
\label{tab:sim_param_B}
\end{table}

\paragraph{Hyperparameters settings.}
%\label{app:hyper}
All the experiments were made with the following hyperparameters. Concerning the soft spectral clustering initialization, the $k$-means algorithm (on the rows of the column leading eigenvectors matrix) is run with 100 initializations. The tolerance threshold $\epsilon$ used to stop the fixed point and the \texttt{VEM} algorithm is set to $10^{-6}$. The maximum numbers of iterations for the fixed point and the \texttt{VEM} algorithm were set to $U_{\max}=50$ and $T_{\max}=50$, respectively.

\subsection{Additional synthetic results}
\label{app:synth_plus}
In this section, we provide additional synthetic results and comments. 

\begin{figure}[htbp]
    \centering
    \includegraphics[width=0.6\textwidth]{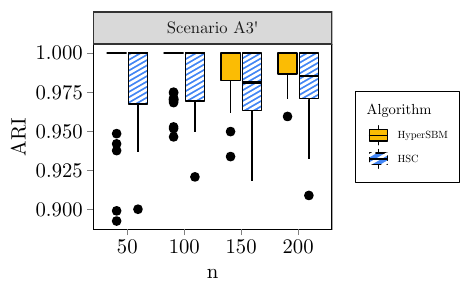}
    \caption{Boxplots of Adjusted Rand Indexes for setting A3', for different number of nodes $n$ (along $x$-axis) and with 2 methods: our \texttt{HyperSBM} (left boxplot) and HSC (right boxplot).}
    \label{fig_SM:ari_res}
\end{figure}

\paragraph{Additional scenario A3'.}
First, we explore clustering  estimation in scenario A3'.  Figure~\ref{fig_SM:ari_res} shows boxplots of the ARI for both our method \texttt{HyperSBM} and HSC. We observe that while \texttt{HyperSBM}  again outperforms HSC (obtaining higher ARI values overall and significantly lower variances), both methods  obtain very good results in this setting (with ARI always larger than 0.9). Thus this setting, while being sparse (see values of $E^{(m,n)}$ in Table~\ref{tab:sim_param_A}) appears to be an easy one from the clustering point of view. This is why we chose to push the limits and introduced setting A3 that is \emph{sparser than the sparse scenario A3'}. Indeed, the ratio between the number of hyperegdes $E^{(n,m)}$ in setting A3' and A3 is equal to 2.7.  In this much sparser regime that we call \emph{highly sparse} below, we have shown that clustering still works with \texttt{HyperSBM} while it fails with HSC. 
%535/198=2.7 ; 229/85=2.7 ; 1080/401= 2.7

\vspace{.5cm}
\vspace{.5cm}

However, setting A3 appears to be too sparse for model selection. Indeed, Table~\ref{tabSM:mod_sel} shows the results of model selection in that setting, where we can see that our ICL criterion performs badly: we always under-estimate the number of groups. It is out of the scope of the present work to further explore the reasons for that nor the  characterization of highly sparse wrt sparse regimes. 
At this point, we conclude that ICL performs well in dense settings (data not shown) as well as sparse ones, but might not in highly sparse regimes. 

\begin{table}[htbp]
\caption{Frequency (as a percentage) of the selected number of groups $Q$ for scenario A3' (true number of groups $Q=3$).  Model selection is carried out by means of the ICL criterion. Results are computed over 50 samples for each value of $n$.} 
	\centering
	\begin{tabular}{C{1cm}C{3cm}C{3cm}C{3cm}C{3cm}}
		\toprule
		$ Q $   & $n=50$    	& $n=100$		& $n=150$	    & $n=200$	        \\
		\cmidrule{1-1}\cmidrule{2-3}\cmidrule{4-5}
		1		& $ 16\% $		& $ 2\% $       & $ 0\% $		& $ 0\% $           \\
		2		& $ 84\% $	    & $ 90\% $      & $ 66\% $		& $ 54\% $          \\
		3		& $ 0\% $	    & $ 8\% $	    & $ 34\% $	    & $ 46\% $          \\
		4		& $ 0\% $		& $ 0\% $       & $ 0\% $		& $ 0\% $           \\
		5		& $ 0\% $		& $ 0\% $       & $ 0\% $		& $ 0\% $           \\
		\bottomrule
	\end{tabular}
	\label{tabSM:mod_sel}
\end{table}

\begin{figure}[h]
\centering
    \includegraphics[width=\textwidth]{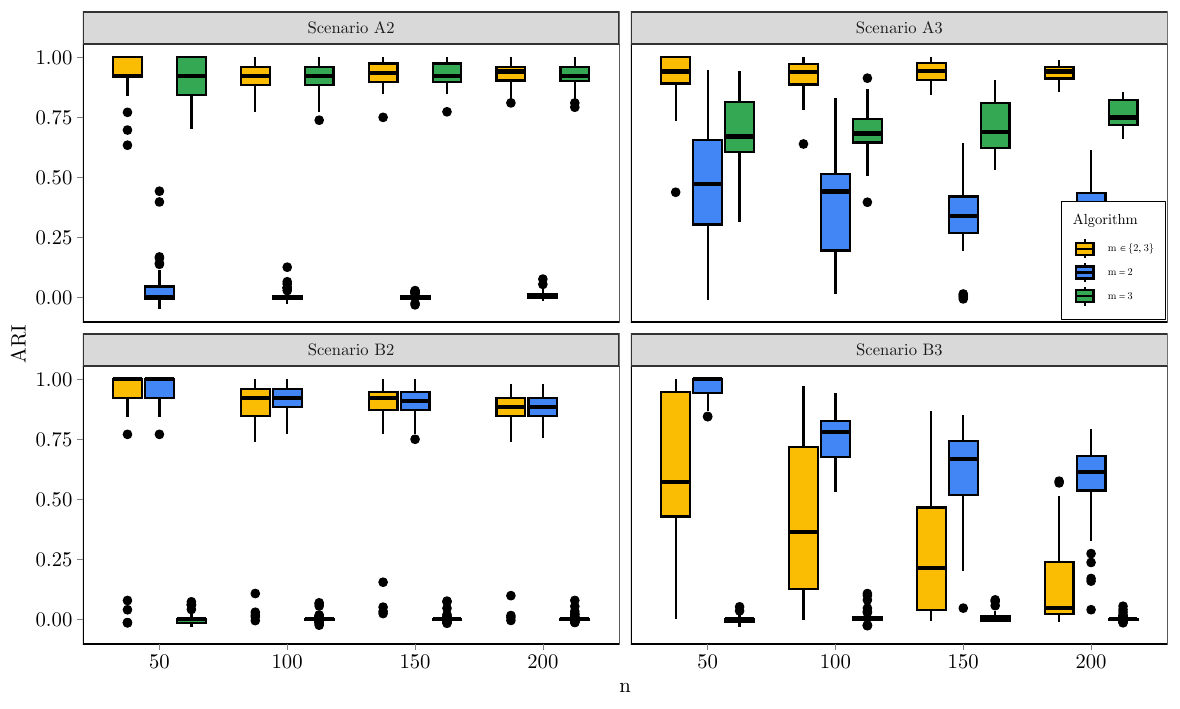}
    \caption{Boxplots of Adjusted Rand Indexes for settings A2, A3, B2 and B3, for different number of nodes $n$ (along the $x$-axis) and with \texttt{HyperSBM} applied on 3 different types of data: the original hypergraphs ($m \in \{2,3\}$, left boxplot), using only size-2 hyperedges ($m=2$, center boxplot) and using only size-3 hyperedges ($m=3$, right boxplot).}
    \label{fig_SM:2or3}
\end{figure}

%%%%%%
\paragraph*{Using only $m$-uniform hypergraphs.} 
\blue{As an additional experiment,  we ran our \texttt{HyperSBM} algorithm on the same datasets but  relying i) only hyperedges of size 2; ii) only hyperedges of size 3. 
%Note that in this way, the hypergraphs become even sparser. 
The results on ARI appear in Figure~\ref{fig_SM:2or3}; we do not provide Mean Squared Relative Error plots here, as the estimated parameters are different among the three models. We obtain that in all scenarios but B3, relying only on size-2 or only on size-3 hyperedges to reconstruct the clusters performs either worse or similarly than with the original non-uniform hypergraph. However, in the dis-assortative  scenario B3, relying only on the size-2 hyperedges improves the clustering. This is not surprising as in this case, there are $Q=3$ different clusters. Observing only size-2 hyperedges, the algorithm learns that any time there is an edge between 2 nodes, they most likely belong to different clusters. Now coming to size-3 hyperedges, the signal is less clear: any time a hyperedge occurs, the algorithm learns that the corresponding nodes most likely either belong to 3 different clusters or that 2 of these nodes can be in one cluster and the last one in another cluster. This situation is more difficult to entangle and the clustering quality degrades. Note however that as the statistician does not have the knowledge that the dataset comes from a dis-assortative scenario with more than 2 groups, she cannot decide to drop the information of the size-3 hyperedges to improve the clustering. 
}

\begin{figure}
     \centering
 \includegraphics[width=0.8\textwidth]{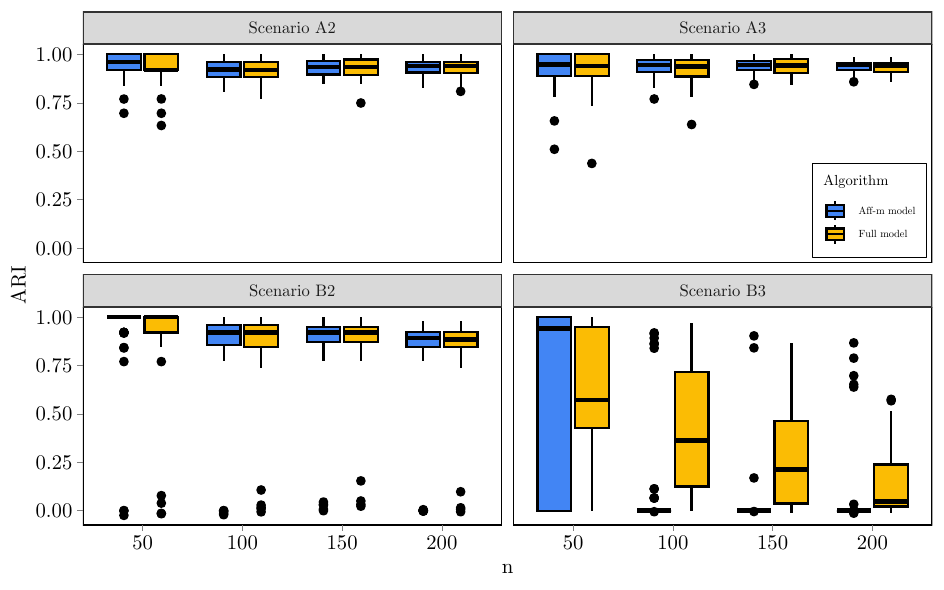}
     \caption{Boxplots of Adjusted Rand Indexes for settings A2, A3, B2 and B3, for different number of nodes $n$ (along the $x$-axis) and with 2 versions of our \texttt{HyperSBM} algorithm: relying on on the (\textbf{Aff-m}) submodel (left boxplot) or on the full HSBM (right boxplot). }
     \label{fig:ARI_affm_vs_full}
 \end{figure}

\begin{figure}
     \centering
 \includegraphics[width=0.8\textwidth]{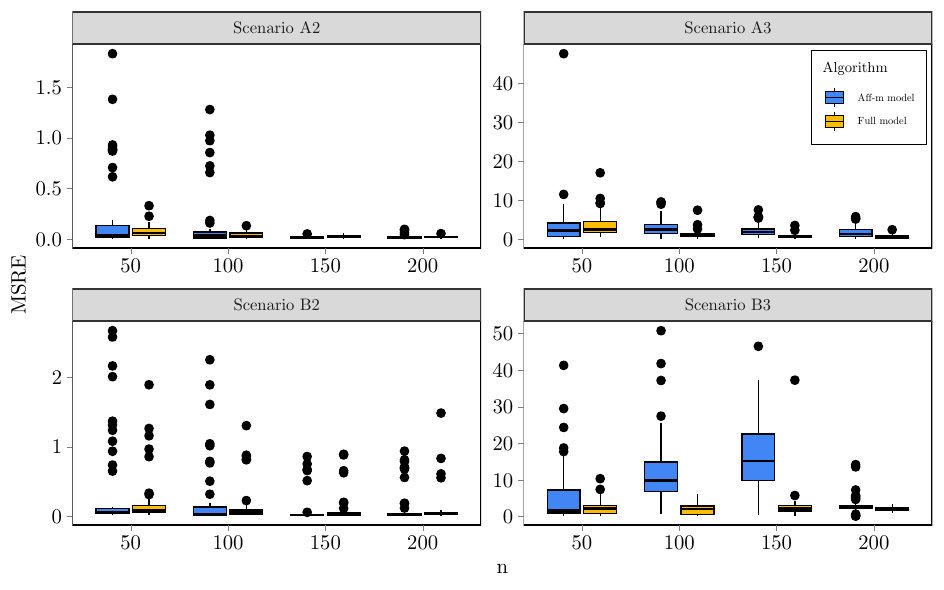}
     \caption{Boxplots of Mean Squared Relative Error for settings A2, A3, B2 and B3, for different number of nodes $n$ (along the $x$-axis) and with 2 versions of our \texttt{HyperSBM} algorithm: relying on on the (\textbf{Aff-m}) submodel (left boxplot) or on the full HSBM (right boxplot). }
     \label{fig:MSRE_affm_vs_full}
 \end{figure}

%%%%%
\paragraph{Using the specific (\textbf{Aff-m}) submodel.}
\blue{
Given that our synthetic datasets were generated under the (\textbf{Aff-m}) submodel, we here compare the version of our algorithm restricted to this case with its full version (whose results were presented in the main manuscript). 
Figure~\ref{fig:ARI_affm_vs_full} shows the boxplots of the corresponding ARIs, while Figure~\ref{fig:MSRE_affm_vs_full} shows the Mean Squared Relative Error (MSRE) of the estimated parameters. 
 We observe that in all scenarios but B3, the results are similar for both versions of our algorithm. In scenario B3 (disassortative with $Q=3$ groups), both the ARIs and MSREs become very bad for the (\textbf{Aff-m}) version of the algorithm. We already observed that this setting is more difficult and apparently, inferring an unconstrained model seems to be a slightly better strategy in this case. }

\begin{figure}
     \centering
 \includegraphics[width=0.8\textwidth]{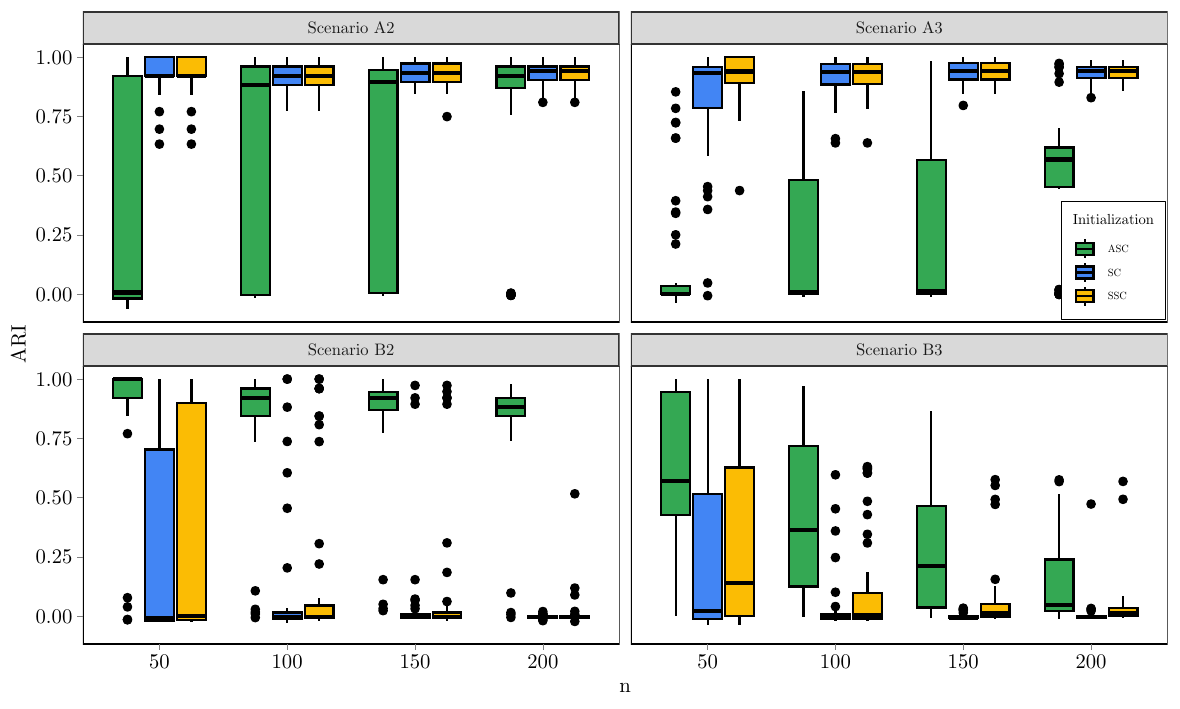}
     \caption{Boxplots of Adjusted Rand Indexes for settings A2, A3, B2 and B3, for different number of nodes $n$ (along the $x$-axis) and with 3 different initializations of our \texttt{HyperSBM} algorithm: relying on spectral clustering (SC, left boxplot), on Soft Spectral Clustering (SSC, center boxplot) or on graph-component absolute spectral clustering  (ASC, right boxplot). }
     \label{fig:ari_Init}
 \end{figure}

\begin{figure}
     \centering
 \includegraphics[width=0.8\textwidth]{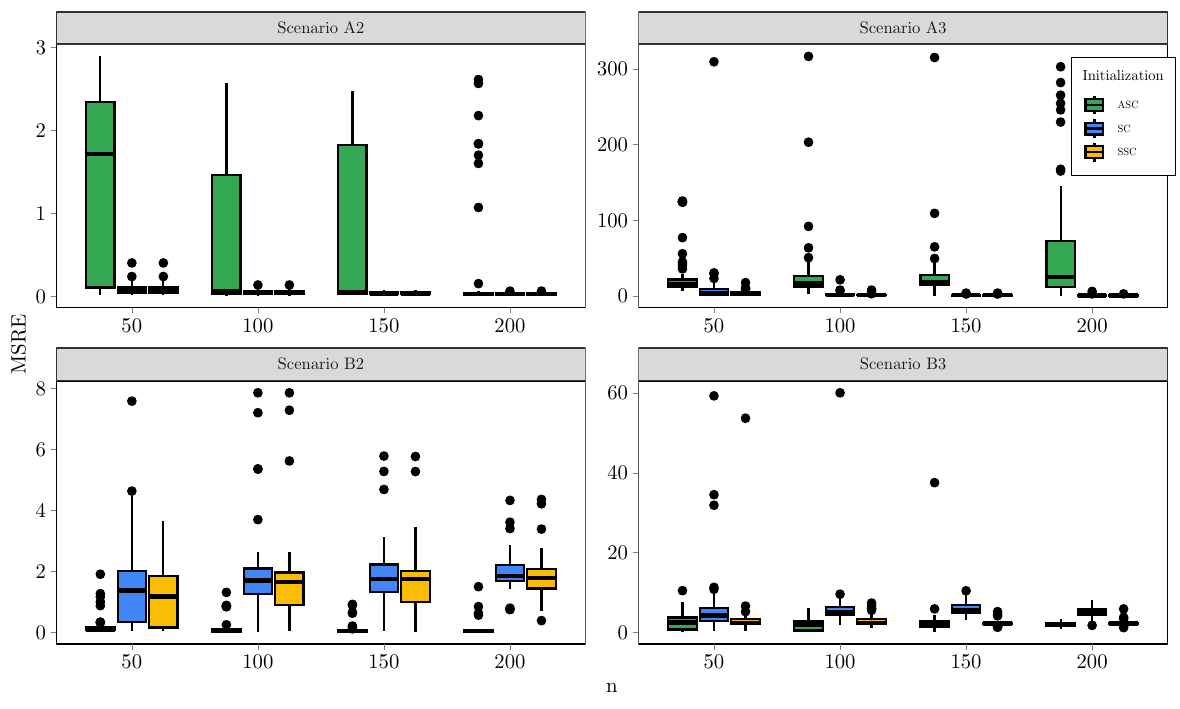}
     \caption{Boxplots of Mean Squared Relative Error  for settings A2, A3, B2 and B3, for different number of nodes $n$ (along the $x$-axis) and with 3 different initializations of our \texttt{HyperSBM} algorithm: relying on spectral clustering (SC, left boxplot), on  Soft Spectral Clustering (SSC, center boxplot) or on graph-component absolute spectral clustering  (ASC, right boxplot). }
     \label{fig:msre_Init}
 \end{figure}

%%%%%%%%%%%%
\paragraph*{Using different initializations.}
\blue{In this experiment, we explore the impact of the initialization choice on the results of our procedure. Let us recall that the results presented in the main manuscript rely on soft spectral clustering (for Scenario A) and graph-component absolute spectral clustering (for Scenario B) initializations (see paragraph ``Algorithm initialization''  in the main manuscript). 
We here compare 3 different initializations: spectral clustering (SC), soft spectral clustering (SCC) and graph-component absolute spectral clustering (ASC) in the 4 settings A2, A3, B2, B3. Figures~\ref{fig:ari_Init} and Figure~\ref{fig:msre_Init} present the results of the boxplots of the ARI and the MSRE in these different settings, respectively. We observe that while SC and SSC give rather similar results, ASC behaves differently. More precisely and as expected, in the assortative scenarios SC and SSC are better than ASC, while in the dis-assortative scenarios B, the converse happens. When analyzing a dataset, as the scenario is not known a priori,  we recommend using many different strategies and select the one with highest optimized ELBO.
}

%%%%%%%%%%%
\section{Analyses on the co-authorship dataset}	
\label{sec_SM:dataset}

The original dataset has 274 papers and 314 authors, with 1 paper having 6 authors and 1 paper having 5 authors. We decided to consider $M=4$ and discard these 2 papers with more than 5 authors. Then, we looked at the largest connected component of the resulting graph. It resulted in 76 papers and 79 authors. 

\begin{figure}[h]
\centering
    \includegraphics[]{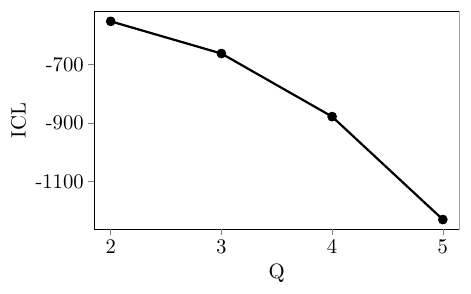}
    \caption{Integrated Classification Likelihood index resulting from fitting the HSBM to the co-authorship dataset with number of latent groups ranging from 2 to 5.}
    \label{fig_SM:ICL}
\end{figure}

We ran \texttt{HyperSBM} with $Q$ ranging from $2$ to $5$, with 2 different initialisations: 1 random and 1 relying on the soft spectral clustering. The random initialisation always gave the best result. The results were robust to different tries. ICL selected $Q=2$ groups, as shown in Figure~\ref{fig_SM:ICL}.

We obtained a first small group with only 8 authors (the remaining 71 authors being in the second large group). 
Inspecting more closely the variational parameters $\tau_{iq}$ for all the nodes, we found that a total of 4 nodes could be considered as  ambiguously classified, while all other nodes had posterior probabilities to belong to one of the group larger than 0.8. More precisely, in the first small group, 2 nodes had posterior probabilities to belong to that group equal to 0.54 and 0.63, respectively; while in the second large group, 2 nodes had posterior probabilities to belong to that group equal to 0.56 and 0.72, respectively.

We discussed in the main text the number of co-authors and degrees in the bipartite graph (i.e. number of co-published papers) of the first small group of authors. We noticed that the 2 authors in this group that had smallest number of co-authors (namely 4) and smallest number of degrees (also 4) are the ones that are ambiguously clustered in this group. While the 2 other authors ambiguously clustered in the second large group have a  number of co-authors of 6 and 4, respectively; and both a  degree of 4. This reinforces the conclusion that on this dataset, \texttt{HyperSBM} has grouped apart the authors which are both the most collaborative and the most prolific ones. 
\\

Then we ran the spectral clustering algorithm on our dataset. 
We looked at the spectral gap, that indicated 15 groups but the gap is not clear. 
Then we looked at the clustering obtained with $Q=2$ groups. 
Spectral clustering output groups with sizes 24 and 55, respectively. We recall that  spectral clustering tends to output comparable sizes groups. 
The small group contains the only author with 12 co-authors and the remaining authors have a number of co-authors ranging from 1 to 4. The second large group has a distribution of the number of co-authors ranging from 1 to 11. 
The small group contains authors with small degree in the bipartite graph, i.e having few  co-published papers (all but one author have degrees less 4 and a last author has degree 7), while the second large group contains the 3 authors with largest degree, the rest of the authors having degrees ranging from 1 to 6. 
%%%
Thus, these groups are neither characterized by the number of co-authors nor by their degrees in the bipartite graph. \\

Finally, we analyzed the same dataset as a bipartite graph under a Bipartite SBM. We relied on the \texttt{R} package \texttt{SBM} through the function \texttt{estimateBipartiteSBM} \citep{SBM_SM}.

The \texttt{Bipartite-SBM} also selected 2 groups of authors (and one group of papers). There was one small group with 4 authors, 
which are exactly the ones that have the highest degree in the bipartite graph and also correspond to the 4 authors having the highest number of co-authors. 

Here, 2 nodes could be considered as ambiguously classified: one node from the first small (resp. second large) group had posterior probability to belong to that group of 0.73 only (resp. 0.67 only).
These 2 nodes where not ambiguously classified by \texttt{HyperSBM} and both appeared in our first small group.

It is interesting to compare the situation of three   particular authors here. 
Author with index 48 has 7 co-authors (the 6th highest) and 6 co-authored papers (the 5th highest). It is outside the small first  group with \texttt{Bipartite-SBM} method (posterior probability $1-0.67=0.33$ to belong to that group); while \texttt{HyperSBM} clusters it unambiguously in the first small group. 
Similarly, author with index 27 has 12 coauthors (1st highest) and only 7 co-authored papers (the 4th highest). This node was ambiguously classified by \texttt{Bipartite-SBM} method in the first small group (posterior probability 0.73 only); while \texttt{HyperSBM} clusters it unambiguously in the first small group. 
Now, conversely, author with index 35 has 8 co-authors (the 6th highest) and 5 co-authored papers (also the 5th highest). This author is unambiguously clustered from the two methods; but while \texttt{HyperSBM} puts it in the first small graph, \texttt{Bipartite-SBM} excludes it from that group. 
The examination of these 3 particular tangent cases seem to show that on this dataset, \texttt{Bipartite-SBM} was more sensible to authors's degrees in the bipartite graph while \texttt{HyperSBM} paid more attention to the sizes of the hyperedges (i.e. number of co-authors) an author was involved in

We also looked at estimated connection probabilities in the bipartite SBM. The authors from the first small group of \texttt{Bipartite-SBM} have many papers (estimated connection probability with the unique group pf papers in the bipartite graph is 11.5\% whereas only 2.5\% for the other large group).  
Finally, we computed the parameters values $\Bqm$ obtained with the groups estimated by \texttt{Bipartite-SBM}. We obtained with $m=2$ that 
$\hat B^{(2)}_{11} \simeq 16,6\%$ (to be compared with $4.2\%$ in \texttt{HyperSBM}); while  $\hat B^{(2)}_{12}\simeq 7\%$ and  $\hat B^{(2)}_{22}\simeq 1\%$ (more similar to the results of \texttt{HyperSBM}, which are $5.1\%$ and $0.8\%$, respectively).  
In this case, the first group of authors behaves differently with respect to within-group connections compared to between-group connections.

\end{document}